\documentclass[11pt]{article}
\usepackage{hyperref}
\usepackage{amsmath}
\usepackage{amssymb}
\renewcommand{\title}[1]{%
    \bigskip%
    \begin{center}%
    \Large\bf #1%
    \end{center}%
    \vskip .2in}

\renewcommand{\author}[1]{%
    {\begin{center}
    #1
    \end{center}}}
\newcommand{\address}[1]{\vspace{-1.7em}\vspace{0pt}
    {\begin{center}
    \it #1
    \end{center}}}

\begin{document}


\title{New approach to nonrelativistic diffeomorphism invariance and its applications}

\author
{
Rabin Banerjee  $\,^{\rm a,b}$,
Pradip Mukherjee $\,^{\rm c,d}$}
\address{$^{\rm a}$S. N. Bose National Centre 
for Basic Sciences, JD Block, Sector III, Salt Lake City, Kolkata -700 098, India }

\address{$^{\rm c}$Department of Physics, Barasat Government College,\\Barasat, West Bengal

 }

\address{$^{\rm b}$\tt rabin@bose.res.in}
\address{$^{\rm d}$\tt mukhpradip@gmail.com}

\begin{abstract}
A  comprehensive account of a new structured algorithm for obtaining nonrelativistic diffeomorphism invariances in both space and spacetime by
 gauging the Galilean symmetry in a generic nonrelativistic field theoretical model is provided. 
 Various applications like the obtention of nonrelativistic diffeomorphism invariance, the introduction of Chern-Simons term and its role in fractional quantum Hall effect, induction of diffeomorphism in irrotational fluid model, abstraction of Newton-Cartan geometry and the emergence of Horava-Lifshitz gravity are discussed in details.
\end{abstract}

\section{Introduction}
Recently, there has been renewed interest in non-relativistic diffeomorphism invariance (NRDI), both from physical and mathematical aspects. This interest was triggered by the papers \cite{SW},\cite{HS} where the role of NRDI to analyze the motion of two dimensional trapped electrons, which is directly connected with the theory of fractional quantum Hall effect (FQHE), was discussed. The relevant field theories involve some variant of the Schr\"odinger theory on a 3-d manifold with universal time. Interestingly, this effective field theory description of FQHE \cite{Son}-\cite{Gromov} found an alternative interpretation in terms of the Newton-Cartan geometry \cite{Son}, \cite{Wu}. However, this introduction of NRDI was adhoc \footnote{As the authors themselves admitted \cite{SW}} i.e. not  following from any systematic prescription or from any geometrical background either. The shortcomings are manifestated in various ways. Thus we find that the space-time transformations become non canonical when one considers time dependent spatial diffeomorphism. It seems that time-dependent and time independent diffeomorphisms are completely disparate and cannot be presented in a unified manner A more serious discrepancy occurs when attempting to recover Galilean symmetry back in the flat limit. We find, surprisingly, that a certain relation between the $ U(1)$ gauge parameter and the Galilean boost parameter must hold \cite{SW}.

NRDI has certain distinct features that sets it apart from usual (i.e. relativistic) diffeomorphism invariance that is so fundamental in understanding the metric formulation of Einstein gravity. For Einstein gravity, the vielbein formulation is related directly with the metric formulation because the spacetime manifold is endowed with a nondegenerate metric. In case of the Galilean space and universal time there is no such structure. Space is relative but time is absolute. In this context it is useful to recall that, following the footsteps of Einstein gravity, space time formulation of Newtonian gravity was worked out by  Elie Cartan \cite{Cartan-1923, Cartan-1924} and subsequently developed by many stalwarts \cite{Havas} -\cite{MALA}. 
The current requirement, however, is a formulation based on the vielbeins which will be an analog of the Cartan formulation of Einstein's gravity. One may think that a suitable algorithm may be obtained from relativistic theories by contraction. However, note that the types of theories that are used, particularly in FQHE, require spatial diffeomorphism which is difficult to obtain by taking a non - relativistic limit of some appropriate relativistic theory. Moreover, sometimes such non - relativistic limits are found to be problematic \cite{Wu}.


         A way out follows from the example of an alternative approach to theories of gravitation, which was pioneered by  Utiyama, Kibble and Sciama \cite{Utiyama:1956sy, Kibble:1961ba, sciama, BH}. They utilised the localisation of the Poincare symmetry of a generic field theory. The resulting theory is called Poincare Gauge theory (PGT). The starting point is a matter theory invariant under global Poincare transformations.The invariance is violated when the parameters of the Poincare transformations are localised by making them functions of spacetime. This invariance may be restored by replacing the ordinary derivatives by suitable covariant derivatives. Theories invariant under local Poincare transformations can be identified with diffeomorphism invariant theories in Riemann - Cartan spacetime. 
         
         Inspired by the Utiyama approach we have developed \cite{BMM1, BMM2, BMM3,AM} a field theoretic method of localising the Galilean symmetry of a generic field theory. Geometrical interpretation of such a localisation or gauging gives nonrelativistic diffeomorphism invariant spacetime. 
  Since the development of the method has reached a stage of finality, a comprehensive account of the method is due. This is all the more necessary because the field theoretic method developed by us is fundamentally different from and is capable of yielding results more general than the method of gauging the centrally extended Galilean algebra \cite{B} which is in vogue for quite some time. Also, apart from the existing applications given in \cite{BMM1,BMM2,BMM3}, several new ones have been found. In the present paper we first trace the ontology of the two different methods (gauging the algebra vis -a-vis gauging the symmetry) in the context of relativistic theories. This is useful for making a transition to the non-relativistic theory where we give a detailed account of the field theoretic method developed by us. New applications have been given from the fields of FQHE and geometry, as well as fluid dynamics. Specifically, we focus on the Chern-Simons theories used in FQHE regarding which there are some confusions in the literature \cite{HS,F}. A significant achievement of our method is to elucidate the nonrelativistic origin of the projectable version of the Horava gravity \cite{H} and to show its difference with Newton-Cartan gravity with particular reference to the boost operation. Also, we consider the hydrodynamic version of 
Schr\"oidinger theory which is equivalent to an irrotational isentropic fluid \cite{Mad, BJ}. A new symmetry related to spatial NRDI is discussed.
           
         The Utiyama approach of constructing PGT may be compared with the approach of gauging the Poincare algebra \cite{N}, mimicking the properties of a non-abelian gauge theory. This gauging prescription introduces gauge fields which can be utilised to define a connection on the tangent bundle. No correspondence can be established between  the gauge field transformations with those due to space-time diffeomorphism at this level. One requires a prescription for connecting the
translation in the tangent space with the general coordinate transformation. This can be done by a constraint on the curvature corresponding to the translation in the group space manifold . The price one has to pay is that the spin connection becomes dependent on the vielbeins and the torsion vanishes.
Thus, connection with relativistic diffeomorphism is established in the perspective of Riemannian space - time. Note that when PGT is obtained by gauging the {\it{Poincare symmetry}} of a generic field theory the Riemann-Cartan spacetime is obtained naturally with or without torsion. The general result involves torsion but one can easily invoke a torsionless connection by symmetrization. Also, PGT gives a complete prescription of coupling a theory, which had Poincare symmetry in the Minkowski space, with curved spacetime. The procedure of gauging the Poincare algebra falls far short of this.

 It is thus no wonder that eventually the main trends of the theoretical formulation of NRDI should emerge from aspects of gauging the symmetry rather than gauging the algebra. In a series of papers \cite{BMM1, BMM2, BMM3} the present authors have shown  how one can
 obtain non relativistic diffeomorphism invariance by localising the extended Galilean symmetry of a model. The original theory is assumed to have invariance under extended Galilean algebra in flat (eucledian) space with time running universally. We make the transformation parameters space time dependent. Naturally, the theory is no longer invariant. To restore the invariance, partial derivatives of the fields need to be replaced by covariant derivatives. New fields are introduced in the process and these become instrumental for the geometric interpretation. The method is therefore similar in spirit to the Utiyama formulation of  PGT. Of course there are non trivial differences that are ultimately connected with the different roles of time in relativistic and non relativistic theories. The approach developed by us has proved to be readily useful in obtaining the most general form of the spatial diffeomorphism. 

  For the sake of comparison, let us briefly consider the approach of obtaining non relativistic diffeomorphism invariance by gauging the extended Galilean algebra ( Bargmann algebra )\cite{B,A,B1,C1,D1,E1}.Here several curvature constraints were imposed to connect the translation parameter of the non abelian gauge group with the diffeomorphism parameter. As a result, the spin connection ceases to be independent. It is expressed in terms of the vierbein leading to a torsionless theory. It should also be stressed that this is a strictly algebraic approach without reference to any action. The dynamical content of the underlying theory becomes somewhat obscure. Thus this formalism is not particularly suitable to couple Galilee invariant theory with curved space or curved spacetime.

  The paper is organised as follows. In Section 2 the method of gauging (or localising) the spacetime symmetry of a field theory is contrasted with the gauging of the symmetry group taking 
  the Poincare transformations as an example.
   The discussion in this section clearly reveals the generality of the localisation of symmetry approach. In Section 3 we discuss the methodology developed recently by us for localising the (extended) Galilean symmetry of a model. A step by step algorithm is prescribed. The inclusion of the $U(1)$ gauge symmetric model is also considered. We find that the same general method of localisation works. Initially an external gauge field is considered. Making the gauge field dynamical did not pose any problems. In particular, the topological C-S interaction fitted in the general scheme without any loss of invariance. In section 4 we discuss several applications of our methodology. The very important issue of diffeomorphism of space \cite{SW, HS, Son, Wu} is addressed here by taking vanishing time translation. However, the spatial diffeomorhism parameter is in general time dependent though time independent parameters have also been considered in the literature \cite{SW}. The unique feature is that, starting from the same formalism, we can discuss both time independent and time dependent spatial diffeomorphisms. Comparison with other approaches is discussed here. Especially, for time dependent spatial diffomorphism, the difficulties and/ or ambiguities reported in the literature \cite{SW,HS} are easily bypassed. The implication of Chern-Simons theory in FQHE is well known. We show how to systematically incorporate such a term within our formalism. Its application to FQHE is then discussed. The covariant derivative that replaces the ordinary derivative in the localisation process reproduces the Hall viscosity and the Wen-Zee term \cite{F}. Our approach finds an interesting application in the context of an irrotational and isentropic fluid obtained by a change of variables in the Schr\"odinger theory. A new form of spatial NRDI is revealed here. When the complete space and time diffeomorphism invariance is considered, the beauty and versatality of our method is full bloomed. We find that the same general transformations hold good. A $(4\times 4)$ matrix structure evolves which completely describes the nonrelativistic space-time geometry. As applications we consider the construction
of the Newton - Cartan geometry and the formulation of Horava - Lifsitz gravity. All the structures of the Newton-Cartan space are reproduced. Contrary to approaches based on gauging the Bargmann algebra (instead of the symmetry), inclusion of torsion is possible. Our results are valid with or without torsion. For the Horava-Lifshitz gravity, its projectable version is obtained. We are able to define the appropriate metric and the physical variables. Their transformations reproduce the known structures. Our analysis provides an important insight into the difference between Newtonian gravity and Horava-Lifshitz gravity which is related to the presence (or absence) of the boost parameter. Concluding remarks are given in section 5.

\section{Poincare gauge theory of gravity} The history of the Poincare gauge theory (PGT) is long. Utiyama \cite{Utiyama:1956sy} first introduced the idea of gauging the {\it{Poincare symmetry}} of a field theory in  Minkowski space. Later on, the theory was developed by many reseachers \cite{Kibble:1961ba, sciama, BH}. The essence of the procedure is the observation that by gauging the Poincare symmetry in  Minkowski space one arrives at a diffeomorphism (diff) invariant theory in the Riemann - Cartan space. This connection has been recently examined by an element to element transformation under diff and Poincare gauge theory transformations \cite{bgmr} and shown to hold for higher derivative matter theories in general \cite{m}. 

 Though nobody was doubtful about the Utiyama procedure, some physisists were not happy to identify the diff parameter as a combination of the translation and rotation parameters of the Poincare group. As a result a diferent approach to PGT emerged. 
This approach is more similar to the one first introduced
by Stelle and West \cite{Nardeli[9]} for the $SO(3,2)$ group spontaneously
broken to the Lorentz group, successively reexamined
by Pagels \cite{Nardeli[10]} for the O(5) group and also used by Kawai
\cite{Nardeli[11]} following the lines of the standard geometrical formulation
of gauge theories. In this framework, one considers
 the Poincare gauge theory as closely as possible
to any ordinary non-Abelian gauge theory, without discarding
the translational part of the Poincare symmetry
in favor of general coordinate transformations. However, this purely algebraic approach is found to be inadequate \cite{N}
and an extra Poincare translation vector has to be introduced. Although the criticisms of the Utiyama approach have been consistently refuted \cite{BH} and the method has found wide applicability [4], the algebraic approach still remains important, particularly in the context of NRDI \cite{B}.

Since our path to the construction of nonrelativistic diffeomorhism invariance (NRDI) rests heavily on the localisation of galilean symmetry it will be appropriate to review both the approaches in the beginning.

\subsection{Lie algebraic approach to Poincare gauge theory} 
The Poincare group is a ten parameter group. Four of them $(\zeta^a, a=0, 1, 2,3)$ refer to translations and six to Lorentz transformations $(\lambda^{ab},$ antisymmetric $\lambda^{ab}=-\lambda^{ba})$. The corresponding generators are $P_a$ and $M_{ab}$, respectively. These generators satisfy the following algebra,
\begin{align}
\label{PA}
\left[ P_a, P_b\right] &=0\notag\\\left[ M_{ab}, P_c\right] &=\eta_{ac}P_b-\eta_{bc}P_a\notag\\\left[M_{ab},M_{cd}\right] &= \eta_{ac}M_{bd}-\eta_{ad}M_{bc}+\eta_{bd}M_{ac}-\eta_{bc}M_{ad}
\end{align}
A global symmetry under the Poincare group means symmetry with constant $\zeta ^a$ and $\lambda^{ab}$. If the parameters are now considered as functions of spacetime, the global symmetry is converted to a local symmetry. New gauge fields are introduced in the process. These are associated with the gauge degrees of freedom and their transformations are worked out in the framework of nonabelian gauge theory. A connection with general coordinate invariance is established via the translation parameter $\zeta ^a$. Different techniques have been adopted to find this connection. We will try to highlight the essence of the problem in our own way.

A Lie algebra valued gauge potential is introduced,
\begin{equation}
\label{PGTF}
A_{\mu}= P_a e_{\mu}{}^a +\frac{1}{2}M_{ab}\omega_{\mu}^{ab}
\end{equation}
The gauge field $e_{\mu}{}^a$is associated with translations while the gauge field $\omega_{\mu}^{ab}$ is associated with Lorentz transformations. Under the usual non-abelian gauge transformations the potential $A_{\mu}$ transforms as,
\begin{equation}
\delta A_{\mu}=D_{\mu}\Lambda=\partial_{\mu}\Lambda+[A_{\mu}, \Lambda]
\label{av}
\end{equation}
where the gauge parameter $\Lambda$ is expressed in terms of the Poincare group parameters as,
\begin{align}
\Lambda=\zeta^a P_a +\frac{1}{2}\lambda^{ab}M_{ab}
\label{la}
\end{align}
analogous to (\ref{PGTF}). From (\ref{PGTF}), (\ref{av}) and (\ref{la}) the transformation rules for $e_{\mu}{}^a$ and $\omega_{\mu}^{ab}$ are obtained by exploiting the Poincare algebra (\ref{PA}).
 \begin{align}
\label{GTV}
\delta e_{\mu}^a &=\partial_{\mu}\zeta^a-\omega_{\mu}{}^a{}_b\zeta^b+\lambda^a{}_b e_{\mu}^b\notag\\\delta\omega_{\mu}^{ab} &=\partial_{\mu}\lambda^{ab}+\lambda^a{}_e \omega_{\mu}^{eb}+\lambda^b{}_e \omega_{\mu}^{ae}
\end{align} 
The Lie algebra valued field strength $F_{\mu\nu}$ which transforms covariantly under the non-abelian gauge transformation is defined as,
\begin{align}
F_{\mu\nu} &=[D_{\mu}, D_{\nu}]=[\partial_{\mu}+A_{\mu}, \partial_{\nu}+A_{\nu}]\notag\\&=P_a F^a{}_{\mu\nu} +\frac{1}{2}M_{ab}F^{ab}{}_{\mu\nu}
\label{Fmunu}
\end{align}
where 
 \begin{align}
F_{\mu\nu}{}^a &=\partial_{\mu}e_{\nu}^a-\partial_{\nu}e_{\mu}^a-\omega_{\mu}{}^a{}_c e_{\nu}{}^c
+\omega_{\nu}{}^a{}_c e_{\mu}{}^c\notag\\F^{ab}{}_{\mu\nu} &=\partial_{\mu}\omega^{ab}{}_{\nu}-\partial_{\nu}\omega^{ab}{}_{\mu}-\omega_{\mu}{}^a{}_c {\omega^c}_{\nu}{}^b+\omega_{\nu}{}^a{}_c 
{\omega^c}_{\mu}{}^b
\label{F}
\end{align}
Our objective is to relate the transformations (\ref{GTV}) with those appropriate to spacetime diffeomorphism and local Lorentz transformations. 
 A close look at the gauge transformations reveal that the variation $\delta\omega_{\mu}^{ab}$
is entirely determined by the local Lorentz rotations whereas translation and rotations are both involved in the transformation of $e_{\mu}^a$. This suggests the possibility that the connection with diffeomorphism can be made through the translation parameter $\zeta^d$ in this gauge approach. For this we define the diffeomorphism parameter,
\begin{equation}
\label{CON}
\xi^\lambda = e^{\lambda}{}_d \zeta^{d}
\end{equation}
The 'geometrical object' $e_{\mu}^a$ will be assumed to be invertible 
\begin{equation}
\label{IV}
e_{\lambda}{}^d e^{\lambda}_{{}e} = \delta^d_e \hspace{.3cm};\hspace{.3cm}e_{\lambda}{}^d e^{\mu}_{{}d}= \delta_\lambda^\mu
\end{equation} 
and is a candidate for the vielbein that transforms the flat Minkowski spacetime to the curved spacetime. However we still have to show that it satisfies the correct transformation rules under general coordinate transformations.

Looking back at the gauge transformations (\ref{GTV}) we can easily see that the transformation of $e_{\lambda}{}^d$ contains  
$\omega_{\mu}{}^a{}_b$ whereas the contrary is not true. The connection proposed between the diff. and translation gauge parameters (see equation (\ref{CON})) indicates the possibility of relating $e_{\lambda}{}^d$ and $\omega_{\mu}{}^a{}_b$ to preserve the internal consistency of the construction.. 
This relation is established by the translation part of the field tensor
$F_{\mu\nu}{}^a$ (see (\ref{F}))
This tensor is a curvature in the gauge space. If we impose the curvature constraint \cite{B}
\begin{equation}
\label{CONSTRAINT}
F_{\mu\nu}{}^a=0
\end{equation}
then $\omega_{\mu}{}^a{}_c$ can be solved in terms of $e_{\nu}{}^c$.
That (\ref{CONSTRAINT})
 is allowed by the local gauge transformations (\ref{GTV}) can be checked directly from (\ref{F}).


To solve for $\omega_{\mu}^{ab}$ we multiply (\ref{F}), subject to (\ref{CONSTRAINT}), by $e^{\mu}_d e^{\nu}_b$ on both sides. This gives
\begin{equation}
0 = e^{\mu}_{d} e^{\nu}_{b} \partial_{\mu} e^{a}_{\nu} - e^{\mu}_{d} e^{\nu}_{b} \partial_{\nu} e^{a}_{\mu} - w_{\mu \phantom{a} b} ^{\phantom{\mu} a} e^{\mu}_{d}  + w_{\mu \phantom{a} d} ^{\phantom{\mu} a} e^{\mu}_{b} 
\label{1}
\end{equation}
Changing $d,\, b$ and $a$ cyclically, we obtain
\begin{equation}
0 = e^{\mu}_{b} e^{\nu}_{a} \partial_{\mu} e^{d}_{\nu} - e^{\mu}_{b} e^{\nu}_{a} \partial_{\nu} e^{d}_{\mu} - w_{\mu \phantom{a} a} ^{\phantom{\mu} d} e^{\mu}_{b}  + w_{\mu \phantom{a} b} ^{\phantom{\mu} d} e^{\mu}_{a} 
\label{2}
\end{equation}
and
\begin{equation}
0= e^{\mu}_{a} e^{\nu}_{d} \partial_{\mu} e^{b}_{\nu} - e^{\mu}_{a} e^{\nu}_{d} \partial_{\nu} e^{b}_{\mu} - w_{\mu \phantom{a} d} ^{\phantom{\mu} b} e^{\mu}_{a}  + w_{\mu \phantom{a} a} ^{\phantom{\mu} b} e^{\mu}_{d} 
\label{3}
\end{equation}
Subtracting (\ref{3}) from the sum of (\ref{1}) and (\ref{2}), we obtain after a few steps
\begin{align}
\label{omega}
\omega_{\mu}^{\phantom{\mu} a b} & = \frac{1}{2} \left[ - e^{\lambda a} \left( \partial_{\mu} e_{\lambda}^{b} - \partial_{\lambda} e_{\mu}^{b} \right) +  e^{\lambda b} \left( \partial_{\mu} e_{\lambda}^{a} - \partial_{\lambda} e_{\mu}^{a} \right) \right.  \notag \\
& \qquad \qquad \qquad \qquad \qquad \left. + e_{\mu}^{c} e^{\lambda a} e^{\rho b} \left( \partial_{\lambda} e_{\rho}^{c} - \partial_{\rho} e_{\lambda}^{c} \right) \right] 
\end{align}
Substitution of $\omega^{ab}_{\mu}$ in $\delta e_{\mu}{}^{a}$ should give the appropriate diffeomorphism transformation of $e_{\mu}^{a}$. To follow the derivation clearly, we will simplify the second term of the right hand side of the first equation of (\ref{GTV}) separately. After some steps we get from (\ref{omega}),
\begin{align}
\omega_{\mu}^{\phantom{\mu} a b} \zeta^b &=\partial_{\mu}\zeta^a-\xi^{\lambda}\partial_{\lambda}e_{\mu}{}^a-\partial_{\mu}\xi^{\lambda}e_{\lambda}{}^a\notag\\
&+\frac{1}{2}\left[\xi^{\lambda}\partial_{\mu}e_{\lambda}{}^a- \xi^{\lambda}\partial_{\lambda}e_{\mu}^a+\zeta^b e^{\lambda b}\partial_{\mu}e_{\lambda}{}^{ a}+\zeta^b e^{\lambda a}\partial_{\lambda}e_{\mu}{}^b +\zeta^b e _\mu{}^c\partial_{\rho}e_{\lambda}{}^{ c} (e^{\rho a}e^{\lambda b} -e^{\lambda a}e^{\rho b} )\right]
\end{align}
Note that if we substitute this in (\ref{GTV}) we find,
\begin{align}
\delta e_{\mu}^{a} &= \xi^{\lambda}\partial_{\lambda}e_{\mu}{}^a + \partial_{\mu}\xi^{\lambda}e_{\lambda}{}^a + \lambda^{a b} e_{\mu}^{b}\notag\\
&+\frac{\zeta^b}{2}\left[\left(e^{\lambda a}\partial_{\lambda}e_{\mu}{}^{b}+ e_{\lambda}^{{}b}\partial_{\mu}e^{\lambda a}\right)+\left( e^{\lambda b}\partial_{\lambda}e_{\mu}^{{}a}+e_{\lambda}^{{}a}\partial_{\mu}e^{\lambda b}\right)  - e _\mu^c e^{\lambda a}\left(e^{\rho b}\partial_\rho e_{\lambda}{}^c + e_{\rho}{}^c\partial_\lambda e^{\rho b}\right)\right]
\label{deltaen}
\end{align}

The expected transformation is reproduced modulo the term in the parenthesis. This term however, does not vanish by algebraic means. One way to obtain the correct transformation relations is to invoke flat geometry in the tangent space and introduce the basis vectors $e_{(a)}$ along with the basis one forms $\omega^{(a)}$. Then the Lie derivative of $\omega^{(a)}$ along $e_{(b)}$ must vanish. Thus, in the coordinate bases \cite{Schutz}
\begin{equation}
e^{\lambda a}\partial_\lambda e_{\mu}{}^ b + e_{\lambda}{}^b\partial_\mu e^{\lambda a} = 0\label{lie}
\end{equation}
This ensures that $e_{\mu}^{a}$ transforms correctly under diff. Two points are to be noted,
\begin{enumerate}
\item $\omega_{\mu}{}^{ab}$ is determined in terms of the vielbein. From the vielbein postulate it can be shown that this leads to a symmetric christoffel connection i.e. torsionless geometry. This is a limitation of the approach. Remember that the Utiyama method leads to Riemann - Cartan geometry with both curvature and torsion.

\item Only setting the gauge curvature to zero does not ensure identification of the translation gauge field with the vielbein. One needs a geometrical input \footnote{The necessity of the geometrical input was understood much earlier \cite{N}}. We have assumed that the Poincare group of transformations act in a Minkowski space which can be identified with the tangent space at a point in the curved spacetime. This gives a direct connection with the Utiyama approach.
\end{enumerate}  
\subsection{Connection with gauging the Poincare symmetry}

The method elaborated above is usually known as a gauging of the Poincare algebra. It is a strictly algebraic approach. Curved space objects like the vierbein or the metric may be appropriately defined, subject to certain restrictions. It is difficult to obtain a dynamical insight through this approach. Indeed it is not clear what ramifications occur if the Poincare symmetry transformations are gauged. It might be recalled that the invariance of an action is checked by considering the Poincare symmetry transformations, hence they are important for dynamical considerations. Below we carry out a simple exercise that illuminates a connection between the algebraic and dynamical approaches.

The infinitesimal form of the usual (global) Poincare transformations are. \begin{equation}
\delta q^a=\lambda^a{}_b q^b+\zeta^a
\label{X}
\end{equation}
If the parameters $\zeta^a$ and $\lambda^a{}_b$ are functions of space-time, then an action previously invariant under (\ref{X}) would no longer preserve that feature. This happens due to the presence of derivatives in the action. In order to restore the invariance under the local transformations, the ordinary derivatives have to be replaced by suitable covariant derivatives. We now observe that two types of such derivatives may be defined. The first one $D_{\mu}q^a$ is such that it transforms inhomogeneously like $q^a$ itself. It is given by,
\begin{equation}
D_{\mu}q^a=\partial_{\mu}q^a+Q^{ab}{}_{\mu}q_b
\label{Y}
\end{equation}
and satisfies the transformation (\ref{X}), 
\begin{equation}
\delta(D_{\mu}q^a)=\lambda^a{}_b D_{\mu}q^b+D_{\mu}\zeta^a
\label{z}
\end{equation}
Here $Q^{ab}{}_{\mu}$ is a new gauge field. The rule (\ref{z}) is obtained provided $Q^{ab}{}_{\mu}$ transforms as,
\begin{equation}
\delta Q^{ab}{}_{\mu}=\partial_{\mu}\lambda^{ab}+\lambda^a{}_e Q^{eb}{}_{\mu}+\lambda^b{}_e Q^{ae}{}_{\mu}
\label{P12}
\end{equation}
It is clear that the new field $Q^{ab}{}_{\mu}$ may be identified with $\omega^{ab}{}_{\mu}$ introduced earlier since both have the same transformation properties (see (\ref{GTV})).
      
We may however recall that restoration of the invariance of the action under gauging happens provided the covariant derivative transforms homogeneously. Thus we define another covariant derivative, 
\begin{equation}
{\cal{D}}_{\mu}q^a=D_{\mu}q^a+Q^a_{\mu}
\label{D}
\end{equation}
such that,
\begin{equation}
\delta({\cal{D}}_{\mu}q^a)=\lambda^a{}_b {\cal{D}}_{\mu}q^b
\label{D1}
\end{equation}
where $Q^a{}_{\mu}$ is a new gauge field like $Q^{ab}{}_{\mu}$ introduced in (\ref{Y}). The rule (\ref{D1}) is satisfied provided,
\begin{equation}
\delta Q^a{}_{\mu}=\partial_{\mu}\zeta^a-{Q_{\mu}}^a{}_b \zeta^b+\lambda^a{}_b Q^b{}_{\mu}
\end{equation}
where use has been made of (\ref{P12}). It is now easy to observe that the new gauge field $Q^a{}_{\mu}$ may be identified with $e^a{}_{\mu}$ used previously. Both have same transformation properties.

\subsection{Gauging the Poincare symmetry}
In discussing Poincare symmetry we have to consider infinitesimal transformations of both fields $\phi(x)$ and the coordinates $x^{\mu}$. It is thus useful to consider two distinct variations. Form variations `$\delta_0$' that change the functional form at the same coordinates, $\delta_0 \phi(x)=\phi'(x)-\phi(x)$ and total variations which account for changes in both the functional form and the coordinates, $\delta \phi(x)=\phi'(x')-\phi(x)$. 

Consider a flat Minkowski space in any dimensions with metric $\eta_{\mu\nu}$. The Poincare group generators are composed of the angular momentum $L_{\mu\nu}=-x_{\mu}\partial_{\nu}+x_{\nu}\partial_{\mu}$, the spin $\Sigma_{\mu\nu}$ whose representation depends on the species of the field being acted upon and the translations $P_{\mu}=-\partial_{\mu}$. The first two are generally expressed in a combined form as $M_{\mu\nu}=L_{\mu\nu}+\Sigma_{\mu\nu}$, which is the total angular momentum.

Under the Poincare group, both coordinates and fields are transformed. The coordinates transform as,
\begin{equation}
\delta_0 x^{\mu}=\left( \frac{1}{2}\theta^{\lambda\nu}L_{\lambda\nu} - \epsilon^\nu P_{\nu}\right) x^{\mu}=\theta^\mu_{\ \nu}x^\nu + \epsilon^\mu=\xi^{\mu}\label{delx}
\end{equation}
while the fields as,
\begin{equation}
\delta_0\phi = -\left(\frac{1}{2}\theta^{\lambda\nu}M_{\lambda\nu}-\epsilon^{\nu}P_{\nu}\right) \phi=\left(-\frac{1}{2}\theta^{\lambda\nu}\Sigma_{\lambda\nu} + \xi^\nu P_\nu\right) \phi\label{delp}
\end{equation}
The total variation of the fields is given by,
\begin{align}
\delta \phi&
= \delta_0\phi+\delta x^{\mu}\partial_{\mu}\phi\notag\\
&= -\frac{1}{2}\theta^{\mu\nu}\Sigma_{\mu\nu}\phi
\label{tp}
\end{align}
obtained on using (\ref{delx}) and (\ref{delp}). Here $\theta^{\mu\nu} (=-\theta^{\nu\mu})$ and   $\epsilon^{\mu}$ are the infinitesimal parameters corresponding to Lorentz transformations and translations, respectively.

Now observe that while the derivatives commute with the form variations, $\delta_0\partial_{\mu}=\partial_{\mu}\delta_0$, they fail for the total variations. Using (\ref{tp}) we find,
\begin{align}
\delta \partial_{\mu}\phi&=\delta_0 \partial_{\mu}\phi+\delta x^{\lambda}\partial_{\lambda}\partial_{\mu}\phi\notag\\&=\partial_{\mu}(\delta\phi)-\partial_{\mu}(\delta x^{\lambda})\partial_{\lambda}\phi\notag\\&=-\frac{1}{2}\theta^{\alpha\beta}\Sigma_{\alpha\beta}\partial_{\mu}\phi-\theta^{\lambda\mu}
\partial_{\lambda}\phi\label{tpd}
\end{align}
while, 
\begin{equation}
\delta_0(\partial_{\mu}\phi)=\partial_{\mu}(\delta_0\phi)=-\frac{1}{2}\theta^{\alpha\beta}\Sigma_{\alpha\beta}\partial_{\mu}\phi-\xi^{\nu}\partial_{\nu}\partial_{\mu}
\phi-\theta^{\lambda\mu}\partial_{\lambda}\phi\label{dp}
\end{equation}
For checking the invariance of the action under the Poincare group of transformations we have to take into account the change in the measure under $x^{\mu}\rightarrow x'^{\mu}$ because the coordinates also change. This is given by the jacobian $\frac{\partial(x')}{\partial(x)}\simeq1+\partial_{\mu}\delta x^{\mu}$.

Then the condition of invariance of the action implies that,
\begin{equation}
\bigtriangleup {\cal{L}}=\delta_0 {\cal{L}}+\delta x^{\mu}\partial_{\mu}{\cal{L}}+(\partial_{\mu}\delta x^{\mu}){\cal{L}}\label{varl}
\end{equation}
vanishes, modulo surface terms.

For global Poincare transformations, (\ref{delx}) implies that,
\begin{equation}
\partial_{\mu}\delta x^{\mu}=\partial_{\mu}\xi^{\mu}=0
\label{d}
\end{equation}
so that the invariance condition becomes,
\begin{equation}
\bigtriangleup {\cal{L}}=\delta_0 {\cal{L}}+\xi^{\mu}\partial_{\mu}{\cal{L}}=0
\label{l}
\end{equation}
Let us consider the localization of the Poincare symmetry by making the parameters $\theta^{\mu\nu}$ and $\epsilon^{\mu}$ functions of spacetime. We may separate coordinate and field transformations by choosing $\xi^{\mu}$ in $\delta x^{\mu}=\xi^{\mu}=\theta^\mu{}_{\nu}x^\nu + \epsilon^\mu$ as the independent parameter instead of $\epsilon^{\mu}$. This gives the freedom of considering generalised transformations with $\xi^{\mu}=0$ but having non-zero $\theta^{\mu\nu}$. We next index the field transformations through $\theta^{ij}$ in Latin and coordinate change through $\xi^{\mu}$ in Greek. The localized Poincare transformations are now defined as,
\begin{align}
\delta \phi &=-\frac{1}{2}\theta^{ij}(x)\Sigma_{ij}\phi\notag\\\delta x^{\mu} &=\xi^{\mu}(x)
\end{align}
This segregation and separate notation will help to define local coordinate frame ($x^i$) that support matter fields and the global (possibly curved) coordinates ($x^{\mu}$).

For local parameters, obviously the original invariance  of the action is lost. There are two reasons. First, the condition (\ref{d}) no longer holds. Secondly, the transformation of the field derivatives given in (\ref{dp}) now changes to,
\begin{equation}
\delta_0(\partial_{k}\phi)=-\frac{1}{2}\theta^{ij}\Sigma_{ij}\partial_{k}\phi-\frac{1}{2}(\partial_k \theta^{ij})\Sigma_{ij}\phi-\xi^{\lambda}\partial_
{\lambda}\partial_{k}
\phi-\partial_k\xi^{\lambda}\partial_{\lambda}\phi\label{dpc}
\end{equation}
Poincare gauge theory emerges from the attempt to modify the matter action so as to obtain invariance under the local Poincare transformations. The process of gauging the global Poincare transformations culminates in the replacement of the ordinary derivatives by appropriate covariant derivatives, such that the latter transform as (\ref{dp}). This is a two step process which is reminiscent of the simple model discussed in the earlier subsection. To begin with, the $\theta$ - covariant derivative is introduced which eliminates the $\partial_k \theta^{ij}$ term from (\ref{dpc}),
\begin{equation}
\nabla_{\mu}\phi=\partial_{\mu}\phi+\frac{1}{2}\omega^{ij}_{\ \ \mu}\Sigma_{ij}\phi\label{covdg}
\end{equation}
We require this derivative to transform as,
\begin{equation}
\delta_0(\nabla_\mu\phi) = -\frac{1}{2}\theta^{ij}\Sigma_{ij}\nabla_{\mu}\phi -(\partial_\mu \xi^{\lambda})\nabla_\lambda\phi-\xi^{\lambda}\partial_{\lambda}
\nabla_{\mu}\phi\label{covv}
\end{equation}
which fixes the transformation of the new fields $\omega^{ij}{}_{\mu}$ as,
\begin{align}
\delta_0 \omega^{ij}{}_{\mu} &= \partial_\mu\theta^{ij}-(\partial_\mu\xi^\lambda) \omega^{ij}{}_{\lambda}-\xi^{\lambda}\partial_{\lambda}\omega^{ij}{}_{\mu}
+\theta^{im} {\omega_m}{}^{j}{}_{\mu}- \theta^{jm} 
{\omega_m}{}^{i}{}_{\mu}\label{delo} 
\end{align}
The total variation of $(\nabla_{\mu}\phi)$ is easily obtained from its form variation (\ref{covv}),
\begin{align}
\delta(\nabla_\mu\phi) = \delta_0(\nabla_{\mu}\phi)+ \xi^{\lambda} \partial_{\lambda} \nabla_{\mu} \phi =-\frac{1}{2}\theta^{ij}\Sigma_{ij}\nabla_{\mu}\phi -(\partial_{\mu} \xi^{\lambda} ) \nabla_{\lambda}\phi\label{covv2}
\end{align}
The presence of the last term spoils the covariant transformation for $\nabla_{\mu} \phi$. This is exactly similar to the simple example considered in section 2.1. Proceeding analogously, we have to define another covariant derivative that will transform covariantly. This is unlike the case of usual gauge theory where one covariant derivative suffices. It is a special feature of taking translations along with Lorentz transformations.

The new derivative is defined as,
\begin{align}
\nabla_k\phi=b_k{}^{\mu}\nabla_{\mu}\phi\label{covl}
\end{align}
which is covariant under Poincare transformations, 
\begin{align}
\delta(\nabla_k\phi)=-\frac{1}{2}\theta^{ij}\Sigma_{ij}\nabla_{k}\phi -\theta^i{}_k\nabla_{i}\phi\label{covltv}
\end{align}
Observe that it satisfies the old (global) rule (\ref{tpd}) which is mandatory for establishing invariance under local Poincare transformations.

Exploiting the earlier results (\ref{covv2}, \ref{covl}, \ref{covltv}) the transformation rule for the new fields $b_k{}^{\mu}$ may be obtained as,
\begin{equation}
\delta b_k{}^{\mu} = \theta_{k}{}^i b_i{}^{\mu}+b_k{}^{\lambda}\partial_{\lambda}\xi^{\mu}\label{delb}
\end{equation}
Having achieved the covariance of derivatives, we are now ready to define an invariant lagrangian {\textit{density}} $\tilde{{\cal{L}}}$ so that the action is invariant. For this, we multiply ${\cal{L}}$ by some function of the new fields such that the discrepancy caused by a non-vanishing $\partial_{\mu}\xi^{\mu}$ (see (\ref{varl}), (\ref{d})) is eliminated. A suitable choice is $b=det(b^i{}_{\mu})$ where $b^i{}_{\mu}$ is the inverse of $b_i{}^{\mu}$ defined by $b^i{}_{\mu}b_i{}^{\nu}=\delta^{\mu}_{\nu}$ and $b^i{}_{\mu}b_j{}^{\mu}=\delta^{i}_{j}$. It may readily be checked that 
\begin{equation}
\delta b+(\partial_{\mu}\xi^{\mu})b=0
\label{b}
\end{equation}
All these arguments eventually lead to a general form of the Poincare gauge theory invariant lagrangian as,
\begin{equation}
{\tilde{\mathcal{L}}}=b{\cal{L}}(\phi, \nabla_k\phi)
\end{equation}
It is possible to develop a geometric interpretation of this lagrangian that effectively connects it to a theory of gravity. The pre-factor b represents the measure (i.e. square root of the determinant of the metric) while $\tilde{\cal{L}}$ characterises the lagrangian density. The details are provided in \cite{bgmr}
\section{Non-relativistic diffeomorphism invariance}

Lest the digressions in the above do not obscure one of the main themes of the paper let us remember that it is to present an algorithm to obtain an invariant theory under nonrelativistic diffeomorphisms which will smoothly reach the  Galilean symmetry in the flat limit. This is achieved in space \cite{BMM1, BMM3} as well as in space time 
for the most general coordinate transformations, consistent with the nature of Galileo - Newton concept of relative space and absolute time. The basic methodology is to localise the Galilean symmetry of a nonrelativistic theory in flat space. In other words, it may also be called as  gauging the Galilean symmetry.

  In the last section we have elaborately reviewed and compared the different approaches to gauging relativistic Poincare  symmetries. The methods of gauging the symmetry group and then connecting to curved spacetime have been shown to be limited to zero torsion spacetime. Moreover, being devoid of any dynamical structure, it is of little use in the direct construction of a diffeomorphic system from a known theory with Poincare symmetry in the Minkowski spacetime. This is precisely obtainable in the localisation of symmetry approach discussed in section 2.2. In fact appeal to an underlying dynamical system can be traced in the algebraic approach also, as we have seen.
   
 The above discussion clearly shows that the approach based on gauging the Bargman (i.e.extended Galilean) algebra \cite{B} is not suitable for our purpose. We thus follow the approach of gauging the  Galilean symmetry of nonrelativistic field theories. Of course there will be fundamental differences due to the different concepts of space and time in non relativistic theories as compared with relativistic theories. It will hence be useful to present the method elaborately, so as to be comprehensive. Accordingly, we divide the analysis in several subsections where localisation of Galilean symmetry, inclusion of gauge fields and geometric interpretation leading to the nonrelativistic diffeomorphism  are discussed separately. 
 
 \subsection{Gauging the (extended) Galilean Symmetry}
 The principle of Galilean relativity states that the outcome of the physical experiments does not change if the system is translated, rotated or boosted in space as a whole or the origin of time be shifted. Stated from the point of view of field theory, the action will remain the same under 
 the infinitesimal global Galilean transformations,
  \begin{eqnarray}
x^0 \to  x^0-\epsilon\nonumber\\
x^i \to  x^i + \eta^{i}-v^{i}t\label{globalgalilean}
\end{eqnarray}
 with
\begin{equation}
\eta^i=\epsilon^{i}+ \lambda^{i}{}_{j}x^{j}
\end{equation} 
 The constant prameters $\epsilon$, $\epsilon^{i}$, $\lambda^{ij}$ and $v^{i}$  respectively represent time and space translation, spatial rotations and galilean boosts.
 $\lambda^{ij}$ are antisymmetric under interchange of the indices.
\footnote{Here $0$ as usual represents time index while letters from the middle of the latin alphabet $(i,j,.....)$ represent spatial indices. When required, they will be represented collectively by $\mu\nu$, etc.}. The transformations (\ref{globalgalilean}) can be formally written as
\begin{equation}
  x^\mu \to x^\mu + \xi^\mu\label{globalgalileans}
\end{equation}
with $\xi^\mu$ given by $\xi^{0} =-\epsilon$ and $\xi^{i} = \eta^{i}-v^{i}t $. Note that $\xi_\mu$ can not be treated as independent diffeomorphisms at this stage. The equation (\ref{globalgalileans}) is just a shorthand way of writing the transformations (\ref{globalgalilean}).

Now assume that the action 
 \begin{equation}
S = \int dt d^3 x {\cal{L}}\left(\phi,\partial_t{\phi}, \partial_k{\phi}\right)
\label{action1}
\end{equation} is
invariant under global galilean transformations ({\ref{globalgalilean}})
Here we have also assumed that $\phi$ is a scalar under the galilean transformations. Our model is otherwise general.

In order to understand the mechanism of localisation it will be helpful to see how the theory ({\ref{action1}}) remains invariant under (\ref{globalgalilean}). Under the general coordinate transformation $x^\mu \to x^\mu + \xi^\mu$ where $x^{\mu} \equiv (t,x,y,z)$ and $\xi^{\mu} \equiv (\xi^0,\xi^{i})$, the action (\ref{action1}) changes by,
\begin{equation}
\Delta S = \int dt d^3x \Delta{{\cal{L}}}
\end{equation}
and
\begin{equation}
\Delta {{\cal{L}}} = \delta_0{{\cal{L}}} + \xi^{\mu}\partial_{\mu}{{\cal{L}}}+ \partial_{\mu}\xi^{\mu}{{\cal{L}}}.
\label{formvariation}
\end{equation} 
Here $\delta_0{\cal{L}}$ denotes the form variation of the Lagrangian. Note the formal semblance with (\ref{varl}). However one should not forget that this similarity is a formal one only. This is because of the very different  
structures of $\xi^\mu$.
For invariance we require,
\begin{equation}
\Delta {{\cal{L}}} =\delta_0{{\cal{L}}} + \xi^{\mu}\partial_{\mu}{{\cal{L}}}+ \partial_{\mu}\xi^{\mu}{{\cal{L}}}= 0.
\label{22}
\end{equation}
Under global Galilean transformation (\ref{22}) is ensured by two conditions. 
First, now $\partial_{\mu}\xi^{\mu} = 0$, which may be explicitly checked using (\ref{globalgalilean}). So (\ref{22}) reduces to, 
\begin{equation}
\Delta {{\cal{L}}} = \delta_0{{\cal{L}}} + \xi^{\mu}\partial_{\mu}{{\cal{L}}} = 0
\label{reducedn}
\end{equation}
Next, this reduced condition is satisfied by the specific form variations of the field and its derivatives given below. The transformation of the field is given by
\begin{eqnarray}
\delta_0\phi = -\xi^{\mu}\partial_{\mu}\phi  - imv^{i}x_i \phi\label{Delphi}
\end{eqnarray}
Note that we are considering symmetry under the infinitesmal transformations of the extended Galilean group. The corresponding boost generator is \cite{BJ,BC}
\begin{eqnarray}
K^i = tP^i + m\int d^3x x^i\phi^*\phi\label{K}
\end{eqnarray}
where $P^i$ is the translation generator. For $ m = 0$, the boost generator for usual Galilean transformation is obtained. The infinitesmal transformation of $\phi $ is then given by
\begin{equation}
 \delta_0 \phi = v^i\{\phi(x^i),K^i\}.
\end{equation} 
Exploiting the fact that the momentum canonically conjugate to $\phi$ is $i\phi^*$, we obtain
\begin{equation}
\delta_0\phi = v^it\partial_i\phi - imv^{i}x_i \phi.
\end{equation}
This reproduces (\ref{Delphi}) 
for the particular case of infinitesmal boost. 


Differentiating (\ref{Delphi}) and utilising the commutative property of differentiation and form variation, we get
\begin{eqnarray}
&\delta_0 \partial_k\phi=-\xi^{\mu}\partial_{\mu}(\partial_{k}\phi)-imv^ix_i\partial_{k}\phi-\lambda^{m}{}_{k}\partial_{m}\phi-imv_k\phi\nonumber\\
&\delta_0 \partial_0\phi= -\xi^\mu\partial_{\mu}(\partial_{0}\phi)-imv^i x_i\partial_{0}\phi+v^{i}\partial_{i}\phi
\label{delkphin}
\end{eqnarray}
 The variations (\ref{Delphi}) and (\ref{delkphin}) satisfy the condition (\ref{reducedn}) and hence these transformations ensure the invariance of (\ref{action1}). Note that it is the particular form of the transformations that ensures the invariance of the generic theory (\ref{action1}) under (\ref{globalgalilean}).

Understanding the last point is crucial for later discussion. So it will be useful to consider a specific example. We chose the Schrodinger field theory as an example of a nonrelativistic model, where the action is given by
\begin{equation}
S = \int dt  \int d^3x  \left[ \frac{i}{2}\left( \phi^{*}\partial_{0}\phi-\phi\partial_0\phi^{*}\right) -\frac{1}{2m}\partial_k\phi^{*}\partial_k\phi\right].
\label{globalaction} 
\end{equation}
The next step will be to prove that this model is invariant under the global Galilean transformations (\ref{globalgalilean}). The variation of $\mathcal{L}$ under (\ref{globalgalilean}) is,
\begin{align}
{\delta}_0 \mathcal{L} & = {\mathcal{L}}'-\mathcal{L}\nonumber\\& =\left[ \frac{i}{2}(({\delta}_0 {\phi}^{*}) {\partial}_{0}\phi+
{\phi}^{*}({\delta}_{0}{\partial}_{0}\phi)) -\frac{1}{2m}(\delta_{0}\partial_{k}\phi^{*})\partial_{k}\phi\right] +c.c.
\label{c}
\end{align}
Now we analyze the individual terms in (\ref{c}),
\begin{equation}
\frac{i}{2}(\delta_{0}\phi^{*})\partial_{0}\phi=\frac{i}{2}\left( \epsilon\partial_{0}\phi^{*}-\eta^{i}\partial_{i}\phi^{*}+v^{i}\left( x^0\partial_{i}\phi^{*}+imx_i\phi^{*}\right)\right) \partial_{0}\phi
\label{7}
\end{equation}
\begin{equation}
\frac{i}{2}\phi^{*}(\delta_{0}\partial_{t}\phi)=\frac{i}{2}\phi^{*}\left( \epsilon\partial_{0}(\partial_{0}\phi)-\eta^{i}\partial_{i}(\partial_{0}\phi)+\left(v^{i} t\partial_{i}-imv^ix_i\right)\partial_{0}\phi+v^{i}\partial_{i}\phi\right)
\label{8}  
\end{equation}
\begin{eqnarray}
-\frac{1}{2m}(\delta_{0}\partial_{k}\phi^{*})\partial_{k}\phi &=& -\frac{1}{2m}\left[ \epsilon\partial_{0}(\partial_{k}\phi^{*})-\eta^{i}\partial_{i}(\partial_{k}\phi^{*})+\left(v^{i} t\partial_{i}+ imv^ix_i\right)\partial_{k}\phi^{*}+\partial_{k}\eta^{i}\partial_{i}\phi^{*}\right.\nonumber\\
&+&\left. imv^{k}\phi^{*}\right] \partial_{k}\phi
\label{9}
\end{eqnarray}
From (\ref{c}) and using (\ref{7}, \ref{8}, \ref{9}) we find that the independent transformation parameters all appear in the expression of the quantity ${\delta}_0 \mathcal{L}$. So the explicit calculation can be performed taking the parameters non zero one at a time. If we take the rotation parameter non zero, the contribution is (taking $\zeta^i = {\lambda^i}_lx^l$)
\begin{eqnarray}
\delta_{0}\mathcal{L}&=&\frac{i}{2} [-\zeta^{i} \partial_{i}\phi^{*}\partial_{0}\phi- \zeta^{i}\phi \partial_{i}\partial_{0}\phi^{*} +\zeta^{i} \partial_{i}\phi\partial_{0}\phi^{*}- \zeta^{i}\phi^{*} \partial_{i}\partial_{t}\phi-\frac{1}{2m}\left[ \zeta^i\partial_{i} \partial_{k}\phi^{*}\partial_{k}\phi +\zeta^i\partial_{i}\partial_{k}\phi \partial_{k}\phi^{*}\right]\nonumber\\
&=&\frac{i}{2} [-\zeta^i\partial_{i}\left(\phi^{*}\partial_{0}\phi -\phi\partial_{0}\phi^{*} \right)] -\frac{1}{2m}[\zeta^i \partial_{i}\left(\partial_{k}\phi \partial_{k}\phi^{*}\right)\nonumber\\
&=& -\zeta^i\partial_{i}\mathcal{L}
\label{akar}
\end{eqnarray}
Similarly we can treat the other parameters. In particular, let us consider the boost related calculations because, apart from the invariance issue, it will clarify another aspect.
For the contribution of the boost part of the global Galilean transformation, (\ref{c}) gives (using \ref{7}, \ref{8}, \ref{9}),
\begin{align}
\delta_{0}\mathcal{L} &=\frac{i}{2} [v^{i} t\partial_{i}\phi^{*}\partial_{0}\phi+ imv^{i}x_i\phi^{*}\partial_{0}\phi +\phi^{*}v^{i} t\partial_{i}\partial_{0}\phi-imv^{i}x_{i}\phi^{*}\partial_{0}\phi+\phi^{*}v^{i}\partial_{i}\phi\notag\\&-v^{i}x^0 \partial_{i}\phi\partial_{0}\phi^{*}+imx_{i}v^{i}\phi \partial_{0}\phi^{*}-\phi v^{i} t\partial_{i}\partial_{0}\phi^{*}-imv^{i}x_{i}\phi \partial_{0}\phi^{*}-\phi v^{i}\partial_{i}\phi^{*}]\notag\\&-\frac{1}{2m}\left[ \left(v^{i} t\partial_{i}+ imv^ix_i\right)\partial_{k}\phi^{*}+ imv_{k}\phi^{*}\right] \partial_{k}\phi\notag\\ &-\frac{1}{2m}\left[ \left(v^{i} t\partial_{i}-imv^ix_i\right)\partial_{k}\phi-imv_{k}\phi\right] \partial_{k}\phi^{*}
\label{akar1}
\end{align}
After a little calculation (\ref{akar1}) simplifies to,
\begin{equation}
\delta_{0}\mathcal{L}=v^{i}t\partial_{i}\mathcal{L}
\label{12}
\end{equation}
Note that this denotes the form variation of the Lagrangian. The total variation, in which we are interested, is given by (\ref{formvariation}) where its last term drops out since global transformations are considered. Thus the total variation under boost is given by,

\begin{align}
\triangle {\mathcal{L}} = \delta_0{\mathcal{L}} -
 v^{i}t {\partial}_{i}{\mathcal{L}}  =0
 \label{11}
\end{align}
This shows the invariance of the Schr\"odinger Lagrangian under boost.
The other parts of the global Galilean transformation also render the action invariant in a similar way. So it can be concluded that the action (\ref{globalaction}) is invariant under the global Galilean transformation.

The last example exhibits how the definite forms of the transformations of the first derivatives of the fields (\ref{delkphin}) together ensure global Galilean invariance. Note that here the transformation parameters are constants.
  If instead, the transformation parameters are functions of space and time, the transformations contain local Galilean parameters. Localisation of (\ref{globalgalilean}) will naturally depend on the Galileo-Newton concept of spaceime.
 Since space is relative and time is absolute the time translation may be generalised as some function of time whereas spatial transformations are functions of both time and space \cite{PLP}. This is also consistent with
Cartan's construction of the 4-dimensional spacetime manifold in terms of an affine connection compatible with the temporal flow $t_{\mu}$ and a rank-three spatial metric $h^{\mu\nu}$. Thus the localised form of (\ref{globalgalilean}) is as follows:
\begin{equation}
\xi^{0}=-\epsilon\left(x^0\right),~~~~~~\xi^i = \eta^i\left(x^0, {\bf{r}}\right) - v^i\left(x^0, {\bf{r}}\right)t
\label{localgalilean}
\end{equation}
where $\eta^i = \epsilon^{i}\left(x^0, {\bf{r}}\right)+\lambda^{i}{}_{j}\left(x^0, 
{\bf{r}}\right)x^{j}$.
Note the difference of this localisation with the localisation of Poincare transformation parameters. Thus the localisation of the Galilean symmetry will have difficulties of its own, making it altogether a new theory. This process will eventually yield an algorithm to construct a diffeomorphic theory from a nonrelativistic field theory.
 
Let us first note that the transformations (\ref{localgalilean}) can be identified with Galilean transformations only in the neighberhood of a point. Hence we will introduce local coordinates $x^a$ at each point of space for the local galilean symmetry group. The local basis ${\bf{e}}^a$ corresponding to local coordinates is assumed to be connected with the global basis ${\bf{e}}^k$.
\begin{equation}
{\bf{e}}^a = \delta_k^a{\bf{e}}^k
\label{basis}
\end{equation}
Note that if one basis is orthogonal then the other is also so on account of (\ref{basis}):\footnote{The vectors in the local basis are labeled by the initial Latin alphabets whereas those of the global (coordinate) basis are labeled using Latin alphabets from the middle.
The time coordinate in the local system will be denoted by $\bar{0}$. Collectively, the local coordinates will be denoted by the initial letters of the Greek alphabet (i.e. $\alpha,\beta$ etc.).}
\begin{align}
{\bf{e}}^a.{\bf{e}}_b&={\delta^a}_k {\delta_b}^{l} {\bf{e}}^k.{\bf{e}}_l\notag\\&=\delta^{a}_k {\delta_b}^{l}\delta^{k}_l\notag\\&=\delta^{a}_b
\end{align}
At the first sight the separation of local coordinates from the global one may appear trivial. However without the local coordinates the parameters of the local Galilean transformations could not be referred and the transformations of the fields cannot be asserted.
Note also that, as we proceed, the distinction between local and global coordinates will become nontrivial  .
 Thus the field $\phi$ will be defined in the local frame. The form variation will be given by
 \begin{eqnarray}
\delta_0\phi = -\xi^{\mu}\partial_{\mu}\phi  - imv^{a}x_a \phi
\label{delphi}
 \end{eqnarray}
 Note the indices of the last term which comes from the phase rotation of $\phi$ due to Galilean boost in the local frame.
 
When the galilean transformation parameters are function of space time the partial derivatives $\partial_k\phi$ and $\partial_t\phi$ no longer transform as (\ref{delkphin}). Following the gauge procedure one needs to introduce covariant derivatives which will transform  as(\ref{delkphin}). Our experience with PGT indicates that this covariant derivative has to be constructed in two steps. The first step in the process of localisation is to convert the ordinary derivatives into covariant derivatives with respect to the global coordinates. Let us introduce the gauge fields $B_0$ and $B_k$ such that,
\begin{eqnarray}
D_k\phi=\partial_k\phi+iB_k\phi\nonumber\\
D_0\phi=\partial_0\phi+iB_0\phi \label{firstcov1}
\end{eqnarray}
The gauge fields $B_0$ and $B_k$ correspond to gauging the rotations and galilean boosts.They have the structures,
\begin{eqnarray}
 B_k = \frac{1}{2}B_k^{ab}\omega_{ab} + B_k^{a\bar{0}}\omega_{a}\nonumber\\
 B_0 = \frac{1}{2}B_0^{ab}\omega_{ab} + B_0^{a\bar{0}}\omega_{a}
\label{gaugefields}
\end{eqnarray}
where $\omega_{ab}$ and $\omega_{a}$ are respectively the generators of rotations and Galileo boosts. Since we consider scalar fields only so $\omega_{ab} = 0$.
However an important exception occurs in two space dimensions where the rotation generator is a scalar. This allows coupling with a scalar field, a fact crucial in the study of the FQHE as we will see. For the sake of generality we will keep the form 
({\ref{gaugefields}}).

 Taking the form variations of (\ref{firstcov1}) and using (\ref{delphi}) we get,
\begin{align}
 \delta_0 D_0\phi &=- \xi^\mu \partial_\mu D_0\phi-     \partial_0\xi^\mu D_\mu\phi -imv^ax_a D_0\phi\nonumber \\
&+i\phi\left(\delta_0 B_0 +\xi^\mu\partial_\mu B_0 + \partial_0\xi^\mu B_\mu -m{\dot{v}^a}x_a 
 \right)\nonumber\\               
\delta_0 D_k\phi &= -\xi^\mu \partial_\mu D_k\phi -\partial_k\xi^\mu D_\mu\phi -imv^ax_aD_k\phi 
                  \nonumber \\
&+i\phi\left(\delta_0 B_k +\xi^\mu\partial_\mu B_k + \partial_k\xi^\mu B_\mu -mv_k- m\partial_kv^ax_a\right)
  \label{gaugefieldsvar}
\end{align}
Note that the variations $\delta_0 B_0$ and $\delta_0 B_k$ are undetermined at this stage. They can be chosen so as to make the terms in the parenthesis vanish. But still the variations given by (\ref{gaugefieldsvar}) differ from the required structure (\ref{delkphin}). This is expected as a similar situation happens in the case of the PGT also. In case of PGT we choose the variations of the $A$ fields at this stage. Now PGT is primarily a gauge theory formulated in the Minkowski space which is a (3+1) dimensional manifold with nondegenerate metric. But here we are localising in $R^3 \times R$. So a difference of methodology in gauging is likely as has been mentioned earlier. Thus here we retain the freedom of choosing the transformation rules of $B_0$ and $B_k$ at a later stage.

   The covariant derivatives with respect to the local coordinates are constructed using the intermediate covariant derivatives (\ref{firstcov1}). We start from the covariant derivative with respect to the local spatial coordinates denoted by $\nabla_a$. It is defined by 
\begin{equation}
\nabla_a\phi={\Sigma_a}^{k}D_k\phi.
\label{nab}
\end{equation}
The new fields $\Sigma_{a}{}^{k}(t, {\bf{r}})$ carry two indices, the lower one refers to the local coordinates and the upper one corresponds to the global coordinates. After some algebra we find that,
\begin{align}
 \delta_0 \nabla_a\phi &= -\xi^\mu \partial_\mu(\nabla_a\phi) - imv^bx_b\nabla_a\phi -imv_a\phi- {\lambda_a}^b\nabla_b\phi
\nonumber\\ &+ D_k\phi\left(\delta_0{\Sigma_a}^k +\xi^\mu\partial_\mu{\Sigma}_a^k -{\Sigma_a}^i\partial_i\xi^k +{\lambda_a}^b{\Sigma_b}^k \right)\nonumber\\
&+i\phi\left[{\Sigma_a}^k\left(\delta_0 B_k   +\xi^\mu\partial_\mu B_k+{\partial}_k\xi^i B_i -m\partial_k v^ax_a-m{v}_k\right) +mv_a\right]
  \label{covariantrule1}
\end{align}
 To ensure that the transformation of $\nabla_a\phi$ is of the same form as $\partial_k\phi$ we impose,
\begin{align}
{\delta}_0 B_{k} &=  -\xi^\mu\partial_\mu B_k - {\partial}_k\xi^i B_i +m\partial_k v^ax_a +mv_k -m{\Lambda_k}^b v_b\nonumber\\
\delta_0 {\Sigma_a}^{k} &=-\xi^\mu\partial_\mu {\Sigma}_a^{k}+ {\Sigma_a}^{i}\partial_{i}\xi^{k} -
{\lambda_a}^b{\Sigma_b}^{k}
\label{delth1}
\end{align}
where  ${\Lambda_k}^a$ is the inverse of ${\Sigma_a}^{k}$, 
\begin{equation}
{\Sigma_a}^{k}{\Lambda_l}^a=\delta^{k}_{l}
\label{sl}
\end{equation}
This ensures the cherished transformation of $\nabla_a\phi$ 
\begin{align}
 \delta_0 \nabla_a\phi = -\xi^\mu \partial_\mu(\nabla_a\phi) - imv^bx_b\nabla_a\phi -imv_a\phi- {\lambda_a}^b\nabla_b\phi
 \label{MM}
 \end{align}
It is easy to show that 
\begin{equation}
\delta_0 {\Lambda_k}^a=
 - \xi^{\mu}\partial_{\mu}{\Lambda_k}^{a} - {\Lambda_l}^{a}\partial_{k}\xi^{l} - {\lambda^a}_{c}{\Lambda_k}^{c}
\label{delLamb}
\end{equation}

 The particular form of localised Galilean transformations (\ref{localgalilean}) indicates that there may be an arbitrary scaling along the time arrow. From the form of $\delta_0D_0\phi$ it is evident that the local covariant derivative $\nabla_{\bar{0}}\phi$ will be a combination of $D_0\phi$ and $D_k\phi$. So we propose,
 \begin{equation}
\nabla_{\bar{0}}\phi=\theta(D_0 \phi+\Psi^k D_k\phi)
\label{finalcov}
\end{equation}
$\theta\left(t\right)$ and $\Psi^k\left({\bf{r}},t\right)$ are additional fields, the transformations of which will be chosen along with that of $B_0$ such that $\nabla_{\bar{0}}\phi$  transforms covariantly,
\begin{equation}
\delta_0\nabla_{\bar{0}}\phi=-\xi^\mu \partial_\mu(\nabla_{\bar{0}}\phi) - imv^bx_b\nabla_{\bar{0}}\phi+v^b\nabla_b\phi
\end{equation} 
Calculation leads to the following results,
\begin{align}
\delta_0B_0 &= m{\dot{v}}^ix_i -\xi^\mu
\partial_\mu B_0 -\partial_0\xi^\mu B_\mu +m{\Lambda_k}^a\Psi^k v_a
\nonumber\\
\delta_0\theta &=-\theta\dot{\epsilon}+\epsilon\dot{\theta}\nonumber\\
\delta_0\Psi^k &=-\xi^\mu\partial_\mu{\Psi}^k+\Psi^i\partial_i\xi^k+\dot{\epsilon}\Psi^k-\frac{1}{\theta}v^b{\Sigma}^{k}_{b}+\frac{\partial}{\partial_0}({\eta}^k-x^0v^k)
\label{delth1n}
\end{align}

The first stage of localisation of the Galilean symmetry (\ref{globalgalilean}) is now over. One now substitutes $\partial_0{\phi}$, $\partial_k{\phi}$ by  
 $\nabla_{\bar{0}}\phi$, $\nabla_a{\phi}$ in the Lagrangian 
${\cal{L}}\left(\phi, \partial_0{\phi}, \partial_k{\phi}\right)$. As a result the Lagrangian becomes $
 {{\cal{L}^{\prime}}\left(\phi, \nabla_{\bar{0}}\phi, \nabla_a\phi\right)}$.


Observe that when the Galilean transformations are localised, $\partial_{\mu}\xi^{\mu}\neq 0$, and so the condition for invariance is given by (\ref{22}) instead of (\ref{reducedn}). What has, however, been achieved is (\ref{reducedn}),
\begin{equation}
\delta_0{{\cal{L}}}' + \xi^{\mu}\partial_{\mu}{{\cal{L}}}' = 0\label{A}
\end{equation}
Thus the modified Lagrangian ${{\cal{L^\prime}}}$ is still not invariant under local Galilean transformations. We remember that the factor $\partial_\mu\xi^\mu $ arises from the Jacobean of the coordinate transformations
This suggests a correction factor $\Lambda$ so that the corrected Lagrangian is,
\begin{equation}
{{\cal{L}}} = \Lambda{{\cal{L^\prime}}}.\label{78}
\end{equation}
Using (\ref{A}) and (\ref{78}), we obtain a condition on $\Lambda$ following from (\ref{22}),
\begin{equation}
\delta_0\Lambda + \xi^\mu \partial_\mu\Lambda 
+ \partial_\mu \xi^\mu\Lambda=0.
\label{lambda}
\end{equation}
 The ansatz,
\begin{equation}
\Lambda = \frac{M}{\theta}
\end{equation}
where 
\begin{equation}
 M = det{\Lambda_k}^a
 \label{M}.
\end{equation}
does the trick, as is explicitly shown below,
\begin{align}
\delta_0\Lambda + \xi^\mu \partial_\mu\Lambda 
+ \partial_\mu \xi^\mu\Lambda &=\left( \frac{\partial\Lambda}{\partial\theta}\delta_0\theta+\frac{\partial\Lambda}{\partial M}\delta_0 M\right) -\epsilon\partial_0\left( \frac{M}{\theta}\right)\notag\\ & +(\eta^i-v^i x^0)\partial_i \left( \frac{M}{\theta}\right) +\left( -\dot{\epsilon}+\partial_i(\eta^i-v^i x^0)\right) \left( \frac{M}{\theta}\right) 
\label{lamb}
\end{align}
But the transformation of M is given by,
\begin{equation}
\delta_0 M=-M{\Lambda_k}^{a}\delta_0{\Sigma_a}^{k}
\label{delM}
\end{equation}
The transformation rule of  $\theta$ is obtained from (\ref{delth1n}) and $\delta_0 M$ can be calculated using (\ref{delth1}).

 Analyzing individual terms in (\ref{lamb}) we see,
\begin{equation}
\frac{\partial\Lambda}{\partial\theta}\delta_0\theta=-\frac{1}{\theta^2}(-\theta\dot{\epsilon}+\epsilon\dot{\theta})
\end{equation}
\begin{equation}
\frac{\partial\Lambda}{\partial M}\delta_0 M=(-\frac{1}{\theta} M{\Lambda_k}^{a}\delta_0 {\Sigma_a}^{k})
\end{equation}
\begin{equation}
-\epsilon\partial_t\left( \frac{M}{\theta}\right)= -\epsilon\left(\frac{\dot{M}}{\theta} -\frac{M}{\theta^2}\dot{\theta}\right) 
\end{equation}
\begin{equation}
(\eta^i-v^i t)\partial_i \left( \frac{M}{\theta}\right)=(\eta^i-v^i t)\partial_i M \left(\frac{1}{\theta}\right) =(\eta^i-v^i t)M{\Sigma_a}^{k}\partial_i {\Lambda_k}^{a}\left(\frac{1}{\theta}\right)
\end{equation}
Adding all terms we get back the condition (\ref{lambda}). Hence it is proved that the Lagrangian is invariant under local Galilean transformation. 
  
 We thus derive the rules of localising the Galilean symmetry of a nonrelativistic model. The algorithm is as follows. Introduce local coordinates at each point of 3-d space. The local basis is connected to the coordinate basis by (\ref{basis}). If the original theory is given by the action (\ref{action1}) which is invariant under the global Galilean transformation
\begin{equation}
x^{\mu}\rightarrow x^{\mu}+\xi^{\mu}
\end{equation}
where $\xi^{\mu}$ is defined in (\ref{globalgalilean}), then
\begin{equation}
S = \int dx^{\bar{0}} d^3x \frac{M}{\theta}{\cal{L}}\left(\phi, \nabla_{\bar{0}}\phi, \nabla_a\phi\right)
\label{localaction}
\end{equation}
is invariant under the corresponding local Galilean transformations (\ref{localgalilean}).

To illustrate the algorithm of localising the Galilean symmetry we consider the Schrodinger field theory (\ref{globalaction}) which was shown to be invariant under the global Galilean transformation. The localised version that follows from our algorithm is obtained in two steps. First, the Lagrangian corresponding to (\ref{globalaction}) is modified by substituting the partial derivatives by the corresponding covariant derivatives. Next, the measure is corrected. The modified Lagrangian including the measure is given by, 
\begin{equation}
\mathcal{L}= \Lambda {\cal{L}'}=\frac{M}{\theta} {\cal{L}'}= \frac{M}{\theta}\left[\frac{i}{2}(\phi^{*}\nabla_{\bar{0}}\phi-\phi\nabla_{\bar{0}}\phi^{*}) -\frac{1}{2m}\nabla_a\phi^{*}\nabla_a\phi
\right].
\label{localschaction} 
\end{equation}
Under the local Galilean transformations the modified Lagrangian ${{\cal{L^\prime}}}$ satisfies (\ref{A}) as can be shown by the following calculation. First,
\begin{equation}
\delta_{0}{\mathcal{L}'}=\left[\frac{i}{2}( (\delta_{0}\phi^{*})\nabla_{\bar{0}}\phi+
\phi^{*}(\delta_{0}\nabla_{\bar{0}}\phi))-\frac{1}{2m}(\delta_{0}\nabla_{k}\phi^{*})\nabla_{k}\phi\right]+c.c.
\label{lprime}
\end{equation}
Expanding the individual terms, we get as follows,
\begin{equation}
\frac{i}{2}(\delta_{0}\phi^{*})\nabla_{t}\phi=\frac{i}{2}\left( \epsilon\partial_{t}\phi^{*}-\eta^{i}\partial_{i}\phi^{*}+v^{i}\left( t\partial_{i}\phi^{*}+imx_i\phi^{*}\right)\right) \nabla_{t}\phi
\label{7'}
\end{equation}
\begin{equation}
\frac{i}{2}\phi^{*}(\delta_{0}\nabla_{t}\phi)=\frac{i}{2}\phi^{*}\left( \epsilon\partial_{t}(\nabla_{t}\phi)-\eta^{i}\partial_{i}(\nabla_{t}\phi)+\left(v^{i} t\partial_{i}-imv^ix_i\right)\nabla_{t}\phi+v^{b}\nabla_{b}\phi\right)
\label{8'}  
\end{equation}
\begin{align}
-\frac{1}{2m}(\delta_{0}\nabla_{k}\phi^{*})\nabla_{k}\phi & =-\frac{1}{2m}[ \epsilon\partial_{t}(\nabla_{k}\phi^{*})-\eta^{i}\partial_{i}(\nabla_{k}\phi^{*})+\left(v^{i} t\partial_{i}+ imv^ix_i\right)\nabla_{k}\phi^{*}\notag\\ & -{\lambda^{b}}_a\nabla_{b}\phi^{*}+ imv^{k}\phi^{*}] \nabla_{k}\phi
\label{9'}
\end{align}
Considering only the contribution of the boost part of the local Galilean transformation we obtain, from (\ref{lprime}),
\begin{eqnarray*}
&\delta_{0}\mathcal{L}'=\frac{i}{2} [v^{i} t\partial_{i}\phi^{*}\nabla_{t}\phi+ imv^{i}x_i\phi^{*}\nabla_{t}\phi +\phi^{*}v^{i} t\partial_{i}\nabla_{t}\phi-imv^{i}x_{i}\phi^{*}\nabla_{t}\phi+\phi^{*}v^{b}\nabla_{b}\phi\\&-v^{i}t \partial_{i}\phi\nabla_{t}\phi^{*}+imx_{i}v^{i}\phi \nabla_{t}\phi^{*}-\phi v^{i} t\partial_{i}\nabla_{t}\phi^{*}-imv^{i}x_{i}\phi \nabla_{t}\phi^{*}-\phi v^{b}\nabla_{b}\phi^{*}]\\&-\frac{1}{2m}\left[ \left(v^{i} t\partial_{i}+ imv^ix_i\right)\nabla_{k}\phi^{*}+ imv_{k}\phi^{*}\right] \nabla_{k}\phi\\&-\frac{1}{2m}\left[ \left(v^{i} t\partial_{i}-imv^ix_i\right)\nabla_{k}\phi-imv^{k}\phi\right] \nabla_{k}\phi^{*}\\
\end{eqnarray*}
which simplifies to,
\begin{equation}
\delta_{0}\mathcal{L}'=v^{i}t\partial_{i}\mathcal{L}'
\label{12'}
\end{equation}
Also, for the boost part,
\begin{equation}
{\xi}^{\mu} {\partial}_{\mu}\mathcal{L}'=-v^{i}t{\partial}_{i}\mathcal{L}'
 \label{11'}
\end{equation}
Thus, from ( \ref{12'}, \ref{11'}),
\begin{equation}
\delta_0 \mathcal{L}'+ {\xi}^{\mu} {\partial}_{\mu}\mathcal{L}'=0
\label{jata}
\end{equation}
Inclusion of $\Lambda$ as done in (\ref{78}) now ensures the invariance condition (\ref{22}).
Therefore, our Lagrangian (\ref{localschaction}) is invariant under local galilean transformation. 

A particularly important point about the localisation of the Galilean symmetry is the 'peculiarity' of transformations of different types of fields. While the transformations in spatial rotation can be categorized easily by the representations of the rotation group the behavior under boost does not fall in such categories. Concerning gauge fields in the nonrelativistic case there is a certain amount of confusion in the literature \cite{SW, HS, F}. Thus, for instance, it was reported \cite{SW} that when the Galilean symmetries of a model with U(1) symmetry was modified to a spatially diffeomorphic model, the Galilean boost symmetry could only be retrieved if a certain relation between the  gauge parameter and the boost parameter is assumed. Clearly, such a relationship can hardly be motivated. Again, in $(2+1)$ dimensional nonrelativistic theories the dynamics of the gauge field may be dictated by the topological Chern - Simons (C-S) term. The C-S dynamics is of immense importance from a practical point of view. In this scenario it is remarkable that some authors reported that the C-S dynamics cannot be included as such in the spatially diffeomorphic theory \cite{HS}. Note that there is no general consensus about this result also. Thus the C-S dynamics has been successfully introduced in \cite{F}. However, in this work much labor is expended to justify the coupling of a scalar field with the spin connection. In our view all these confusions can be traced to the lack of a consistent method of coupling a gauge theory with curvature of the space. It is therefore reassuring to know that in our scheme gauge fields are naturally accommodated {\it{as per the same general programme}}   which was used for a scalar field theory. This will be discussed now.

\subsection{Inclusion of gauge field} 
We have explained how global Galilean symmetry is localised taking the simplest example of a scalar field theory. We will now include interaction with a $U(1)$ gauge field, keeping one of the most significant areas of application to the theory of FQHE \cite{SW} in mind. The starting action with global Galilean symmetry is
\begin{equation}
S = \int dx^0 d^2 x {\cal{L}}\left(\phi,\partial_{\mu}{\phi}, A_\mu, \partial_\mu A_{\nu}\right)
\label{action2}
\end{equation}
Apart from the invariance of the action (\ref{action2}) under global Galilean transformations \ref{globalgalilean} it is assumed to be symmetric under local abelian gauge transformations
\begin{align}
\phi&\rightarrow\phi+i\Lambda\phi\nonumber\\A_{\mu}&\rightarrow A_{\mu}-\partial_{\mu}\Lambda
\label{gt}
\end{align}
This additional symmetry poses a challenge to our localisation procedure. The localisation of the Galilean invariance and its subsequent geometric interpretation should not disturb the $U(1)$ gauge invariance of the original model.

 We now discuss the issue of global Galilean transformations of the gauge fields in some details. Under these transformations (\ref{globalgalilean}) the  complex scalar field $\phi$ and its derivatives $ \partial_k\phi$ and $\partial_0\phi $ transform as in equation (\ref{delkphin}). 
The transformations of $A_{\mu}$ and its various derivatives are required as new input.
As we have mentioned earlier, due to the intricacies of non-relativistic spacetime the transformation of various fields (under boost) must be determined
from case to case. The transformations of the gauge potential were obtained in 
\cite{L}. Of course $A_k$ transforms as a vector under rotation while $A_0$ transform as a scalar under the same. Combining these, the transformations of $A_\mu$ under global Galilean transformations are written as
 \begin{align}
\delta_0 A_0 &= \epsilon\partial_0 A_0 - \eta^{l}\partial_l A_0 + tv^{l}\partial_l A_0+v^l A_l\nonumber\\
\delta_0 A_i &= \epsilon\partial_0 A_i - \eta^{l}\partial_l A_i + tv^{l}\partial_l A_i+{\lambda_i}^l A_l
\label{delA}
\end{align}
Then the transformations of their derivatives can be shown to be
\begin{align}
\delta_0 \partial_k A_0 &=\epsilon\partial_{0}(\partial_{k} A_0)-\left(\eta^{l} - x^0v^l\right)\partial_{l}(\partial_{k}A_0)+\lambda_k{}^{l}\partial_{l}A_0+v^l\partial_k A_l\nonumber\\
\delta_0 \partial_0 A_0 &=\epsilon\partial_{0}(\partial_{0}A_0)-\left(\eta^{l} - x^0v^l\right) \partial_{l}(\partial_{0}A_0)+v^l \partial_l A_0+v^{l}\partial_{0}A_l
\label{delA0}
\end{align}
and 
\begin{align}
\delta_0 \partial_k A_i &=\epsilon\partial_{0}(\partial_{k} A_i)-\left(\eta^{l}-x^0v^l\right)\partial_{l}(\partial_{k}A_i) + {\lambda_k}^l \partial_l A_i+\lambda_i{}^{l}\partial_{k}A_l\nonumber\\
\delta_0 \partial_0 A_k &=\epsilon\partial_{0}(\partial_{0}A_k)-\left(\eta^{l}- x^0v^l\right)\partial_{l}(\partial_{0}A_k)+v^{l}\partial_{l}A_k
+\lambda_k{}^{l}\partial_{0}A_l
\label{delAi}
\end{align}
These are the transformations that ensure 
\begin{equation}
\delta_0{{\cal{L}}} + \xi^{\mu}\partial_{\mu}{{\cal{L}}} = 0\label{reduced}
\end{equation}
Also, $\partial_\mu\xi^\mu = 0$. Together they keep $\delta S = 0$  under the global Galilean transformations,
where $S$ is given by (\ref{action2}). 
 
Now we make the transformations local by invoking $\ref{localgalilean})$
One should remember that after localisation these transformations can be viewed as Galilean transformations only locally. The final form of the local Galilean invariant theory will thus refer to the local coordinates. This explains the introduction of the local coordinates $x^a$ (see equation (\ref{basis})), 
notwithstanding the fact that in flat euclidean space they are trivially connected with the global coordinates. 

Once the parameters of the transformations are local
 the partial derivatives of $\phi, A_0, A_i$ with respect to space and time will no longer transform as (\ref{delkphin}, \ref{delA0}, \ref{delAi}). Following the gauge procedure one needs to introduce covariant derivatives which will transform covariantly as (\ref{delkphin}, \ref{delA0}, \ref{delAi}) with respect to the local coordinates. As we have shown above, the first step in the process of localisation is to convert the ordinary derivatives into covariant derivatives with respect to the global coordinates. To begin with, introduce the gauge fields $B_\mu$ to define covariant derivatives of the complex scalar field $\phi$ with respect to space and time in global coordinate as,
\begin{eqnarray}
\tilde{D}_\mu\phi=\partial_\mu\phi+iB_\mu\phi
 \label{firstcovn}
\end{eqnarray}
Similarly new gauge fields $C_\mu, F_\mu$ will be introduced to define global covariant derivatives for  the fields $A_\mu$ as,
\begin{align}
\tilde{D}_\mu A_0 &=\partial_\mu A_0+iC_\mu A_0
\notag\\
\tilde{D}_\mu A_i &=\partial_\mu A_i+iF_\mu A_i
\label{firstcovg}
\end{align}
Note that different sets of gauge fields are introduced for $A_0$ and $A_i$, a typical signature of Galilean spacetime. Also note the structural similarity of the global covariant derivatives in each case.

In the next step the global covariant derivatives are converted to
covariant derivatives with respect to space and time in local coordinates. For the complex scalar field these local covariant derivatives were already defined (\ref{nab})
 We found that the local covariant derivatives transform covariantly (\ref{curvedcov}) provided the
 additional fields transformed as (\ref{delth1}) and (\ref{delth1n}) 

 The new feature of the present model is the inclusion of the gauge fields $A_\mu$ in the original action.
We follow a similar procedure to construct the appropriate local covariant derivatives for these fields.
\begin{align}
\nabla_a A_{\bar{0}} &={\Sigma_a}^{k}\tilde{D}_k A_0\notag\\
\nabla_{\bar{0}} A_{\bar{0}} &=\theta(\tilde{D}_0 A_0+\Psi^k \tilde{D}_k A_0)\notag\\
\nabla_a A_b &=({\Sigma_a}^{k}\tilde{D}_k A_i){\delta^i}_b\notag\\
\nabla_{\bar{0}} A_b &=\theta(\tilde{D}_0 A_i+\Psi^k \tilde{D}_k A_i){\delta^i}_b
\label{nabA}
\end{align}
Plugging the expressions of $\delta_0 {\Sigma_a}^{k}, \delta_0\Psi^k, \delta_0\theta$, the local covariant derivatives will transform as the usual ones (see, equations (\ref{delA0}),(\ref{delAi})) 
\begin{align}
\delta_0(\nabla_a A_{\bar{0}}) &=\epsilon\partial_{0}(\nabla_{a} A_{\bar{0}})-\left(\eta^{l}-v^{l} x^0\right)\partial_{l}(\nabla_{a}A_{\bar{0}})
+\lambda_a{}^{b}\nabla_{b}A_{\bar{0}}+v^b \nabla_a A_b\notag\\\delta_0 (\nabla_{\bar{0}} A_{\bar{0}}) &=\epsilon\partial_{0}(\nabla_{\bar{0}}A_{\bar{0}}) -\left(\eta^{l}-v^{l} x^0\right)\partial_{l}(\nabla_{\bar{0}}A_{\bar{0}})+v^b \nabla_b A_{\bar{0}}+v^{b}\nabla_{\bar{0}}A_b\notag\\
\delta_0 (\nabla_a A_b) &=\epsilon\partial_{0}(\nabla_{a} A_b)-\left(\eta^{l}- v^{l} x^0\right)\partial_{l}(\nabla_{a}A_b)+{\lambda_a}^c \nabla_c A_b+\lambda_b{}^{c}\nabla_{a}A_c\nonumber\\
\delta_0 (\nabla_{\bar{0}} A_b) &=\epsilon\partial_{0}(\nabla_{\bar{0}}A_b)-\left(\eta^{l}-v^{l} x^0\right)\partial_{l}(\nabla_{\bar{0}}A_b)
+v^{a}\nabla_{a}A_b
+\lambda_b{}^{c}\nabla_{\bar{0}}A_c
\label{globalcov}
\end{align}
provided the newly introduced fields transform as,
\begin{align}
\delta_0 C_0 &=\epsilon\dot{C_0}
+\dot{\epsilon}C_0-(\dot{\eta}^i
-\dot{v}^{i}x^0)C_i-({\eta}^i-v^{i}x^0)\partial_i C_0+v^l C_l+i{A_0}^{-1}\dot{v}^l A_l
\nonumber\\
\delta_0 C_k&=\epsilon\dot{C}_k
-\partial_k({\eta}^i-v^{i}x^0)C_i-({\eta}^i-v^{i}x^0)\partial_i C_k+i{A_0}^{-1}\partial_k({v}^l) A_l\nonumber\\
\delta_0 F_0 &=\epsilon\dot{F}_0+\dot{\epsilon}F_0
-(\dot{\eta}^l-\dot{v}^{l}x^0)F_l-({\eta}^l-v^{l}x^0)\partial_l F_0+v^l F_l\notag\\
\delta_0 F_k &=\epsilon\dot{F}_k-\partial_k
({\eta}^l-v^{l}x^0)F_l-({\eta}^l-v^{l}x^0)\partial_l F_k
\end{align}
Certain interesting features in the construction of the local covariant derivatives for the gauge fields are to be noted. First, we assume the same basic structure for constructing the corresponding global covariant derivatives as was done for the complex scalar field earlier. Second, it is remarkable that the same basic fields are employed to convert global to local covariant derivatives with the same set of transformation rules. This explains why these fields are connected with the geometry of the non-relativistic spacetime \cite{BMM2}. Indeed, this is the genesis of the obtention of Newton-Cartan space-time, as elaborated in section 4.8. 

 The localization of Galilean transformation and symmetry restoration for the action is now straightforward. Following the same approach stated above , the action will be modified, replacing the    partial derivatives  by the local covariant derivatives. We will now have to amend for the fact that $\partial_\mu \xi^\mu $ is no longer zero. But we know how to do it. Invoke the  correction factor(\ref{M}) in the measure of integration. This prescription leads to the action
\begin{equation}
S = \int dx^{\bar{0}} d^2x \left(\frac{M}{\theta}\right){\cal{L}}\left(\phi,\nabla_\alpha{\phi},A_\alpha, 
\nabla_\alpha A_\beta\right)
\label{localaction1}
\end{equation}
 where $\alpha,\beta \equiv {\bar{0}, a}$. The action (\ref{localaction1}) is invariant under the local Galilean transformations (\ref{localgalilean}). The structure of the action where the derivatives are replaced by covariant derivatives indicates that the $U(1)$ gauge symmetry of the flat space model (\ref{action2}) is preserved. In the following we will address this issue in connection with specific models.

We will now discuss  various applications of the formalism developed in this section. The keynote is that the theory (\ref{localaction1}) can be reinterpreted as a geometric theory where the connection between the global and the local coordinates will be nontrivial. This will lead naturally to diffeomorphism invariant theories in a nonrelativistic setting.

\section{Applications} As we have repeatedly emphasised, the motivation of the Galilean gauge theory discovered in \cite{BMM1, BMM3} came from the requirements of the theory of FQHE \cite{SW, HS}. In the theory of FQHE trapped electrons moving on a plane are considered. The most general symmetry,that goes beyond the usual Galilean symmetry, in this context, is the spatial diffeomorphism symmetry \cite{SW}. So we begin with the abstraction of spatial diffeomorphism. The approach is valid for general $D$
dimensional space where $D$ will be taken to be two in the study of FQHE. 

\subsection{Emergence of spatial diffeomorphism}
We will now show that our formalism leads to diffeomorphism invariant theory in space. Since the  goal is  diffeomorphism in space we take the time translation in (\ref{localgalilean}) vanishing,
\begin {equation}
\epsilon(x^0)= 0\label{spacediff}
\end {equation} 
 The second equation of (\ref{delth1n}) and (\ref{spacediff}) then show that  $\theta$ = constant. Without any loss of generality it can be taken to be one. 
 The local Galilean transformations (\ref{localgalilean}) with $\epsilon = 0$ is then equivalent to,
\begin{equation}
x^k \to x^k + \xi^k\label{genco}
\end{equation}
where $\xi^k$ is an arbitrary function of space and time defined in (\ref{localgalilean}). To give these the sense of local galilean transformations we have to refer them to the local coordinates. We know how to write the corresponding locally Galilean invariant theory discussed in the last section. The action is 
\begin{equation}
S = \int dx^0 d^2x M{\cal{L}}\left(\phi,\nabla_\alpha{\phi},A_\alpha, 
\nabla_\alpha A_\beta\right)
\label{localaction12}
\end{equation}
which is obtained from the action (\ref{localaction1}) by substituting $\theta = 1$. Also note that
\begin{eqnarray}
\nabla_{\bar{0}}\phi & = & {\Sigma_{\bar{0}}}^0D_0\phi + {\Sigma_{\bar{0}}}^kD_k\phi\nonumber\\
& = & \theta D_0\phi + \theta\Psi^kD_k\phi =  D_0\phi + \Psi^kD_k\phi
\end{eqnarray}

We have interpreted the action (\ref{localaction12}) to be a theory invariant under Galilean transformations in $x^a$
with $\Psi^k$, $\Sigma_a^{{}{k}}$, $B_k$ as new fields that are functions of $x^a$ and $t$. But due to the trivial nature of the transformations (\ref{basis}) they can be as well considered as fields that are functions of global flat coordinates $x^k$ and time.
However the form of the action (\ref{localaction12}) indicates that it can be given a much more elegant interpretation. Forget about the triviality of the local spatial coordinates and elevate it to the status of locally inertial coordinates in the tangent space at $x^k$. In this new interpretation the coordinates labeled by $`a'$ define an orthogonal basis with origin at the point of contact and the local and global coordinates agree on a patch of curved space containing the origin. In this scheme $\Sigma_a^{{}{k}}$ and its inverse act as vielbeins that transform from locally inertial cordinates to global ones and vice versa. 
This indicates the possibility of reinterpreting the invariance of (\ref{localaction1}) under (\ref{localgalilean}) as diffeomorphism invariance under (\ref{genco}) in curved space. The coordinates labeled by $\mu$ define the coordinate basis in the curved space. That this interpretation is consistent will be shown below.

 Let us re-examine the structure of the transformation of ${\Sigma_a}^k$ which is obtained from (\ref{delth1}) under  the condition $\epsilon = 0$ as, 
 \begin{equation}
\delta_0 {\Sigma_a}^{k} = {\Sigma_a}^{i}\partial_{i}
\xi^k - \xi^i\partial_{i}{\Sigma_a}^{k}+
{\lambda_a}^b{\Sigma_b}^{k}
\label{delth11n}
\end{equation}
Note the dual aspects of the  transformation. With respect to the  coordinates $x^i$ it satisfies the transformation rules of a contravariant vector under the general coordinate transformation (\ref{genco}) whereas with respect to the coordinates $x^b$ it is a local rotation. From the transformation of $\Lambda_{k}{}^{a}$ given by (\ref{delLamb}) we find to our delight that it transforms as a covariant vector under diffeomorphism (\ref{genco}) corresponding to its lower tier index $k$ while as an euclidean vector under rotation corresponding to its local index $a$. It will thus be reasonable to propose the following connection between local and global coordinates in the overlapping patch
\begin{equation}
dx_a = {\Sigma_a}^{k}dx_k\label{v}
\end{equation}
and
\begin{equation}
dx^a = {\Lambda_k}^{a}dx^k\label{vv}
\end{equation}
where ${\Lambda_k}^{a}$ is the inverse of ${\Sigma_a}^{k}$. In the above relations (\ref{v}) and (\ref{vv}), $dx^k$ is the differential increment of the general curved space coordinate.
Note that, contrary to (\ref{basis}), the above connections have become nontrivial.

 

 We will next show that we can construct a metric (and its inverse) for the  manifold from the fields ${\Sigma_a}^k$ and its inverse $\Lambda_{k}{}^{a}$. Let us define
\begin{equation}
g_{ij}=\delta_{cd}{\Lambda_i}^c {\Lambda_j}^d
\label{metric1}
\end{equation}
as `the metric'. From the transformation rules for ${\Lambda_i}^c$ we can prove that under (\ref{genco}), $g_{ij}$ transforms as a covariant tensor 
\begin{equation}
\delta_0 g_{ij} =-\xi^k \partial_k g_{ij}-g_{ik}\partial_j\xi^k-g_{kj}\partial_i\xi^k
\label{diff} 
\end{equation}
This butresses the claim of $g_{ij}$ to be a metric. However the most important role of the metric is to give the invariant distance between two points. Here, in local coordinates  
the distance between two points is given by $dx_adx_a$. Using (\ref{v}) and the definition (\ref{metric1})we obtain
\begin{eqnarray}
dx^adx^a &=& {\Lambda_k}^{a}dx^k{\Lambda_l}^{a}dx^l
\nonumber\\
&=& \delta^{ab}{\Lambda_k}^{a}dx^k{\Lambda_l}^{b}dx^l
\nonumber\\
&=& g_{kl}dx^kdx^l                    
\end{eqnarray}
Indeed, $g_{ij}$ is nondegerate as it admits the following inverse
\begin{equation}
g^{kl}=\delta^{ab}{\Sigma_a}^{k}{\Sigma_b}^{l},~~~g^{kl}g_{lm}=\delta^k_m
\label{invmetric}
\end{equation}
which can be checked explicitly.

 Now, following (\ref{metric1}) a scintillating result follows 
\begin{eqnarray}
M = \rm{det}{\Lambda_i}^c = \sqrt{g}
\end{eqnarray} 
We can therefore propose the following replacement, recalling $\theta=1$,
\begin{equation}
\int dx^{\bar{0}} d^2x \frac{M}{\theta} \to \int dx^0 d^2x \sqrt{g}\label{measure}
 \end{equation}
in (\ref{localaction1}).
Note that the action (\ref{localaction1}) is invariant under certain transformations in flat space in the global flat coordinates where the new fields $\theta, \Psi_k, \Sigma_k^{{},a}$
appear. These transformations could be identified with Galilean transformations only in local coordinates which are defined by (\ref{basis}). It was gratifying to observe that the new fields may be interpreted as geometric objects that transform from the global curved space to the local tangent space. The  transformation of the measure given in (\ref{measure})is thus interpreted as change of measure from local inertial coordinates to the coordinates that charts {\it{globally curved space}}. The emergence of the required measure to be compatible with general coordinate transformation (\ref{genco}) is delightful to observe. By this reinterpretation of the fields we get curved geometry. The idea of spatial diffeomorphism that has surfaced in the theory of FQHE \cite{SW, HS} from an empirical point of view is thus shown to have a deep connection with the localisation of Galilean symmetry. In section 4.2 we will elaborate on FQHE.

Now events happen not only in space but at a certain time instant also. Though we are working with vanishing time translation, the appearance of time in the diffeomorphism parameter $\xi$ makes
the time arrow relative at different points of curved space. The time component of the vectors in the local flat coordinates will not be simply equal to the time component of the vectors in the curved space. That is why we have distinguished the corresponding indices from the beginning. To relate the time components we will use the remaining field $\Psi^k$ and its transformation rule from (\ref{delth1n}). Naturally this transformation rule does not show obvious geometric interpretation (spacetime is not a single manifold). However it fits with the concept of emergent spatial diffeomorphism, as we will see. In the following, when we discuss space time invariance both $\theta$ and $\Psi^k$ will play crucial roles. 

From a practical point of view our theory gives a structured algorithm of constructing spatially diffeomorphic theory from Galilean symmetric theories with the general structure of (\ref{action2}). To establish this algorithm we have to see how the transformations of the fields and the covariant derivatives obtained from the gauge approach in the previous section can be written in the backdrop of curved space.

 The local coordinates map the tangent space at a space point. Geometric quantities are defined in the tangent space and allow arbitrary rotations. However since the local system is tied to a point in the curved space, Galilean boost is now no longer included in the local transformations. It is now absorbed in the spatial diffeomorphism. With this picture in mind, the transformations of the physical fields have to be investigated.

The fields $\phi$ and $A_{\bar{0}}, A_a$ are at our disposal. Using equations (\ref{Delphi}) and (\ref{delA}) we can write these rules in the local coordinates as
\begin{eqnarray}
\delta_0\phi &=&  - \xi^{k}\partial_k\phi - imv^{a}x_a \phi\nonumber\\
\delta_0 A_{\bar{0}} &=& - \xi^{k}\partial_k A_{\bar{0}} +v^b A_b\nonumber\\
\delta_0 A_a &=&  - \xi^{k}\partial_k A_a +{\lambda_a}^b A_b
\label{phi2}
\end{eqnarray}
In terms of these we will define the appropriate fields in the curved space.
Remember in this context that this mapping can only be achieved in the overlap of the two systems i.e in the neighborhood of the origin of the local system.


We start with the scalar field $\phi$.
The transformation of the scalar field
in the curved space is obtained from (\ref{phi2}) as 
\begin{equation}
\delta_0\phi = - \xi^i\partial_i\phi\label{phic}
\end{equation}
Note that in the new interpretation the two descriptions match in the neighborhood of the origin of local coordinate system. This is why the last term of the corresponding equation of  (\ref{phi2}) does not appear in (\ref{phic}).

Components of the vector field ${\bf{A}}$ are connected by a relation similar to  (\ref{v}),
\begin{equation}
A_a = {\Sigma_a}^{k}A_k
\label{Ac}
\end{equation}
The transformation of $A_a$ is the Galilean transformation given in (\ref{phi2})
and that of ${\Sigma_a}^{k}$ is given by (\ref{delth11n}). The resulting transformation of $A_k$ in the curved basis is obtained by equating the form variations of both sides of (\ref{Ac}).
A straightforward calculation yields
\begin{equation}
\delta_0 A_k  =  - \xi^{i}\partial_i A_k - \partial_k\xi^iA_i\label{delAicurved}  
\end{equation}
This is the required transformation for a covariant vector. 

Particular care is needed for discussing the time components of the fields. As has been already emphasized, though there is no time translation but time is involved in the spatial diffeomorphism parameters. The time component with respect to the local coordinates (denoted by an overbar on zero)
is to be related to the time component in curved coordinates by the following ansatz
 \begin{equation}
 A_{\bar{0}}  = A_0 + \Psi^{k} A_k \label{delAbar0curved}  
\end{equation}
Exploiting (\ref{phi2}), the transformation rule for $A_0$ is then worked out as 
\begin{equation}
\delta_0 A_0  =  - \xi^{i}\partial_i A_0 - \dot{\xi}^iA_i\label{delA0curved}  
\end{equation}
The structure of the above transformation is to be noted. The second term is dependent on the time variation of the diffeomorphism parameter  which can only be avoided if we consider time independent transformations. The structure of (\ref{delA0curved}) is the paradigm of the transformation of time components in the curved space, as will be subsequently observed.

After obtaining the transformations for the basic fields the geometric interpretation is established on a firm ground. However, the issue of substituting
the covariant derivatives $\nabla_{{\bar{0}}}\phi$, $\nabla_k\phi$, $\nabla_aA_b$, $\nabla_{{\bar{0}}}A_a$, $\nabla_aA_{\bar{0}}$ and $\nabla_{{\bar{0}}}A_{\bar{0}}$ by appropriate derivatives with respect to the curved coordinates still remains. We denote these respectively by $D_0\phi$, $D_k\phi$, $D_kA_l$, $D_0A_l$, $D_kA_0$ and $D_0A_0$. The following definitions are proposed:
\begin{eqnarray}
\nabla_{a}\phi &=& {\Sigma_a}^k D_k\phi\nonumber\\
\nabla_{\bar{0}}\phi &=& D_0\phi + \Psi^kD_k\phi\nonumber\\
\nabla_{a}A_b &=& {\Sigma_a}^k {\Sigma_b}^l D_k A_l\nonumber\\
\nabla_{\bar{0}}A_a &=& {\Sigma_a}^{k}\left(D_0 A_k + \Psi^l D_l A_k\right)\nonumber\\
\nabla_a A_{\bar{0}} &=& {\Sigma_a}^{k}\left(D_k A_0 + \Psi^l D_k A_l\right)\nonumber\\
\nabla_{\bar{0}} A_{\bar{0}} &=& D_0A_0 +\Psi^k D_k A_0 +\Psi^k D_0 A_k +\Psi^k\Psi^l D_kA_l
\label{curvedcov}
\end{eqnarray}
Note that the construction of the time component of the covariant derivatives mimics our prescription (\ref{delAbar0curved}). Furthermore, there is a structural similarity of the above relations with those covariant derivatives defined in the global and local coordinates. For instance, the first relation in (\ref{curvedcov}) matches with (\ref{nab}). However, whereas ${\Sigma_a}^k$ in (\ref{nab}) is just a field, it is a vielbein in (\ref{curvedcov}). Also, there are other subtle differences in interpretation which will slowly unfold.

The transformation laws of the new derivatives in curved space are once again obtained from the transformations rules (\ref{delAicurved}), (\ref{delA0curved}) and (\ref{globalcov}). To illustrate our method we take the transformation of $D_k\phi$ and show the calculation explicitly. Taking the form variation of both sides of the first equation of (\ref{curvedcov}) we get
\begin{equation}
\delta_0\left(\nabla_{a}\phi\right)=
\left(\delta_0{\Sigma_a}^k\right)D_k\phi
+ {\Sigma_a}^k\left(\delta_0D_k\phi\right)
\label{show}
\end{equation}
From (\ref{MM}) we write
\begin{equation}
\delta_0\left(\nabla_{a}\phi\right) =
-\xi^b\partial_b\left(\nabla_{a}
\phi\right) -imv^b\nabla_{a}\left(x_b\phi\right)+\lambda_a{}^{b}\nabla_{b}\phi
\end{equation}
The penultimate term of the above expression will have vanishing contribution because in the overlap of the two coordinate systems, $x_b\phi $ must be smoothly vanishing. Substituting this result on the left hand side of (\ref{show}) and using the transformation of $\Sigma_a{}^{k}$ we get the transformation $\delta_0D_k\phi$.
Working in an analogous way we get the transformation rules of the other curved space derivatives. The results are summarised as 
\begin{eqnarray}
\delta_0 D_k\phi &=& -\xi^i\partial_i\left(D_k\phi \right) - \partial_k\xi^i D_i\phi\nonumber\\
\delta_0 D_0\phi &=& -\xi^i\partial_i\left(D_0\phi \right) - \dot{\xi}^k D_k\phi\nonumber\\
\delta_0 D_k A_l &=& -\xi^i\partial_i\left(D_k A_l \right) - \partial_k\xi^m D_m A_l -\partial_l\xi^m D_k A_m\nonumber\\
\delta_0 D_0 A_k &=& -\xi^i\partial_i\left(D_0 A_k \right) - \partial_k\xi^l D_0 A_l -\dot{\xi}^l D_l A_k\nonumber\\
\delta_0 D_k A_0 &=& -\xi^i\partial_i\left(D_k A_0 \right) - \partial_k\xi^l D_l A_0 -\dot{\xi}^l D_k A_l\nonumber\\
\delta_0 D_0 A_0 &=& -\xi^i\partial_i\left(D_0 A_0 \right) - \dot{\xi}^k\left( D_k A_0 +  D_0 A_k\right)\label{varcurvedcov}
\end{eqnarray}
Note that all the curved space derivatives defined by (\ref{varcurvedcov}) transform canonically, following the transformations corresponding to their component labels established for the field components. For example, the expression for $\delta_0(D_k \phi)$ shows that $D_k\phi$ transforms as $A_k$ (see equation (\ref{delAicurved})). Similarly $D_0\phi$ transforms as $A_0$ (see (\ref{delA0curved})). The higher rank tensors like $D_k A_l$ transform appropriately.

For explicit calculations we will require expressions for the derivatives 
$D_k \phi, D_0\phi, D_k A_l, D_0 A_k, D_k A_0$ 
in terms of the basic fields with well defined transformations. These expressions are obtained by requiring consistency with (\ref{varcurvedcov}). Following this we define the derivatives $D_0\phi$ and $D_k\phi$ as,
\begin{align}
D_0\phi &= \partial_0\phi + i{\cal{B}}_0\phi\nonumber\\
D_k\phi &= \partial_k\phi + i{\cal{B}}_k\phi
\label{dkphi}
\end{align}
where the transformation rules for the newly introduced fields ${\cal{B}}_0$ and ${\cal{B}}_k$ are given by,
\begin{align}
\delta_0 {\cal{B}}_0  &=  - \xi^{i}\partial_i {\cal{B}}_0 - \dot{\xi}^i {\cal{B}}_i\nonumber\\
\delta_0 {\cal{B}}_k  &=  - \xi^{i}\partial_i {\cal{B}}_k - \partial_k\xi^i {\cal{B}}_i\label{delBicurved}  
\end{align}
We observe that ${\cal{B}}_k$ transforms as a covariant spatial vector (see (\ref{delAicurved})) and ${\cal{B}}_0$ transforms in the same way as the time component of vectors are expected to transform in our formalism ( see equation (\ref{delA0curved}). It is also gratifying to see that the transformations ${\cal{B}}_\mu$ as given by (\ref{delBicurved}) is nothing but the transformations of ${B}_\mu$ which were introduced during gauging Galilean symmetry in the limit of spatial diffeomorphism, i.e. no time translation symmetry and exclusion of Galilean boost from the local transformations (See equations (\ref{delth1}) and (\ref{delth1n})). This not only shows the internal consistency of our construction but also indicate a larger geometric implication of the method.
Later, in the following we explore this elaborately.

A word about the introduction of the new field ${\cal{B}}$ is useful. Observe that the set of vector fields $A$ were present in the original model. The new vector fields ${\cal{B}}$ emerge from the localization prescription that leads to our formulation in curved space.

Similarly we define the other derivatives acting on `A's in the following way,
\begin{eqnarray}
 D_i A_k &=& \left(\partial_i A_k - \partial_k A_i\right) + i({\cal{B}}_i A_k-{\cal{B}}_k A_i)\nonumber\\
 D_0 A_k &=& \left(\partial_0 A_k - \partial_k A_0\right) + i({\cal{B}}_0 A_k-{\cal{B}}_k A_0\nonumber\\
 D_k A_0 &=& \left(\partial_k A_0 - \partial_0 A_k\right) + i({\cal{B}}_k A_0-{\cal{B}}_0 A_k)
\label{covexpand}
\end{eqnarray}
such that they satisfy the transformation rules (\ref{varcurvedcov}).

The algorithm for the construction of the spatially diffeomorphic nonrelativistic theories can now be summarised:
\begin{enumerate}
\item We assume a non relativistic matter field theory in space which is Galilean invariant. For simplicity and priority we consider a complex scalar field but otherwise the action is generic. The theory may be endowed with other (internal) symmetries. In this paper we have taken $U(1)$ symmetry as the additional symmetry, obviously keeping an eye for applications to FQHE.  
\item Gauge the Galilean symmetry by replacing the derivatives of the field by the corresponding local covariant derivatives.
Also correct the measure appropriately as in (\ref{localaction1}). The resulting theory is now locally Galilean invariant theory.
\item Take time translation vanishing. The local Galilean transformations are then equivalent to general coordinate transformations in curved space.  
\item Formulate the theory as a theory invariant under general coordinate transformations in a curved space by the substitution (\ref{measure}) and by replacements of the covariant derivatives in the action (\ref{localaction1}) by the covariant derivatives in the curved space. Use the definitions (\ref{curvedcov}).

\item The diffeomorphic theory obtained in the above procedure will contain the fields $\Sigma_{a}{}^{k}$ and $\Psi^k$. The fields $\Sigma_{a}{}^{k}$ will be grouped to give rise to tensors in the curved space e.g the metric tensor. The fields $\Psi^k$ are independent fields in the theory without any kinetic term. Later, in this paper we will relate $\Psi^k$ to the Newton Cartan geometry.
\end{enumerate}
Following this algorithm the spatial diffeomorphishm invariant action corresponding to the Galilean invariant theory (\ref{action2}) is given by,
\begin{equation}
S=\int dx^{\bar{0}} d^2 x \sqrt{-g}\mathcal{L}(\phi, D_{\mu}\phi, A_{\mu}, D_{\mu}A_{\nu})\label{ssl}
\end{equation}

\subsection{Fractional Quantum Hall Effect}
The concept of spatial diffeomorphism invariance finds application in the theory of FQHE \cite{SW, HS}. We therefore start from an example which models a non relativistic  electron moving in an external gauge field,
given by the action, 
\begin{equation}
S = \int dx^0  \int d^2x_k  \left[ \frac{i}{2}\left( \phi^{*}\Delta_{0}\phi-\phi \Delta_0\phi^{*}\right) -\frac{1}{2m}\Delta_k\phi^{*}\Delta_k\phi\right]
\label{globalaction2} 
\end{equation} 
where
\begin{eqnarray}
\Delta_{0}\phi = \partial_0\phi + iA_0\phi\nonumber\\
\Delta_{k}\phi = \partial_k\phi + iA_k\phi
\end{eqnarray}
and $A_\mu$ is the external gauge field. The theory (\ref{globalaction2}) is invariant under global Galilean transformations (\ref{globalgalilean}) as can be checked explicitly. Note that the theory is also invariant under $U(1)$ gauge transformation (\ref{gt}). 

Simplifying (\ref{globalaction2}) we can get,
\begin{align}
S = \int dx^0 \int d^2 x_k &\left[ \frac{i}{2}\left( \phi^{*}\partial_{0}\phi-\phi\partial_0\phi^{*}\right)-\phi^{*}\phi A_0-\frac{1}{2m}\partial_k\phi^{*}\partial_k\phi-\right.\notag\\
& \left.\frac{{A_k}^2}{2m}\phi^{*}\phi+\frac{i}{2m}A_k(\phi^{*}\partial_{k}\phi-\phi\partial_k\phi^{*})
\right]\label{gcs}
\end{align}

 The corresponding theory invariant under local Galilean transformations (\ref{localgalilean}), according to our algorithm,
is
\begin{align}
S = \int dx^{\bar{0}} \int d^2x_a \frac{M}{\theta}&\left[\frac{i}{2}\left(\phi^{*}\nabla_{\bar{0}}\phi-\phi \nabla_{\bar{0}}\phi^{*}\right)-\frac{1}{2m}\nabla_a\phi^{*}\nabla_a\phi -\phi^{*}\phi A_{\bar{0}}-\right.\notag\\&\left.\frac{{A_a}^2}{2m}\phi^{*}\phi+\frac{i}{2m}A_a(\phi^{*}\nabla_{a}\phi-\phi\nabla_a\phi^{*})
\right]
\label{localscintaction} 
\end{align}
In the following we will consider spatial diffeomorphism ($\epsilon = 0$) where $\theta = 1$.
We can then transform our results in a geometric setting following the algorithm presented earlier.
 
Let us first consider the special case when $\xi$, the spatial diffeomorphism parameter, is time independent. From the definition of $\xi^i$, (\ref{localgalilean}), we find $v=0$. Then the third equation of (\ref{delth1n}) shows that, along with the time independence of $\xi$, $\Psi_k = 0$ may be chosen. Under this condition,  $\nabla_{\bar{0}}\phi=D_0\phi$. After some algebra the action (\ref{localscintaction}) reduces to, 
 \begin{align*}
S &= \int dx^0 \int d^2x (det{\Lambda_k}^a)\left[\frac{i}{2}\left(\phi^{*}D_{0}\phi-\phi D_0\phi^{*}\right)-\phi^{*}\phi A_0-{\Sigma_a}^k{\Sigma_a}^l \left(\frac{1}{2m}D_k\phi^{*}D_l\phi\right) -\right.\notag\\&\left.{\Sigma_a}^k{\Sigma_a}^l\left(\frac{1}{2m}A_k A_l\phi^{*}\phi\right)+{\Sigma_a}^k{\Sigma_a}^l\left(\frac{i}{2m}A_k(\phi^{*}D_{l}\phi-\phi D_l\phi^{*})\right)
\right]
\end{align*}
Using the definition of the metric (\ref{invmetric}) this is reduced to a generally covariant theory in the curved space
\begin{align}
S &=\int dx^0 d^2x (det{\Lambda_k}^a)\left[ \frac{i}{2}\left( \phi^{*}(D_{0}+iA_0)\phi-\phi(D_0-iA_0)\phi^{*})\right)\right.\notag\\&\left. -g^{kl}\frac{1}{2m}(D_k-iA_k
)\phi^{*}(D_l+iA_l)\phi\right]
\label{localscaction1} 
\end{align}
The action (\ref{localscaction1}) can now be written as a non-relativistic diffeomorphism invariant action,
\begin{equation}
S = \int dx^0 d^2x \sqrt{g}\left[ \frac{i}{2}\left( \phi^{*}{\bar{D}}_{0}\phi-\phi{\bar{D}}_0\phi^{*}\right) -g^{kl}\frac{1}{2m}\bar{D}_k\phi^{*}\bar{D}_l
\phi\right]\label{diffaction3}
\end{equation}
where 
\begin{eqnarray}
\bar{D}_{0}\phi = D_0 \phi+i A_0\phi=\partial_0\phi +i\left(A_0 + {\cal{B}}_0\right)\phi\nonumber\\
\bar{D}_{k}\phi = D_k\phi+iA_k\phi=\partial_k\phi +i\left(A_k + {\cal{B}}_k\right)\phi
\label{dbar}
\end{eqnarray}

So we can interpret from the result that localization of Galilean symmetry for the non-relativistic field theoretic model of complex scalar fields interacting with a vector field in flat space gives a theory with an action invariant under general coordinate  transformation in curved space. Note that we have considered the spatial diffeomorphism parameter as time independent and there is no time translation.

It may be mentioned that, in the absence of the gauge field, the diffeomorphism invariant action would be given by (\ref{diffaction3}) with $A_{\mu}=0$. The flat space limit of that action would correspond to the theory of complex scalars (\ref{globalaction2}) without any gauge interaction.  
 In the presence of gauge fields the action (\ref{diffaction3}) involves the fields 
  $A_{\mu}$  and ${\cal{B}}_{\mu}$ in the combination $(A_{\mu}+{\cal{B}}_{\mu})$. 
 Also, $A_{\mu}$ and ${\cal{B}}_{\mu}$ have identical transformation properties {\footnote{This is true for general coordinate transformations. If gauge transformations are also included then $A_{\mu}$ and ${\cal{B}}_{\mu}$ have different transformations.}}. Effectively, therefore, by a field redefinition, there is only one field- say $A_{\mu}$ and we may write the diffeomorphism invariant action as (\ref{diffaction3}) with,
\begin{align}
\bar{D}_{0}\phi&=\partial_0\phi +iA_0\phi\notag\\
\bar{D}_{k}\phi&=\partial_k\phi +iA_k\phi\label{P11}
\end{align}
 Indeed this feature (having $A_{\mu}$ and ${\cal{B}}_{\mu}$ in the specific combination ($A_{\mu} + {\cal{B}}_{\mu}$)) is not a general characteristic and would not hold if the gauge field $A_{\mu}$ was dynamical. This has been illustrated (see subsections 4.4 and 4.6) for the specific example where complex scalars were coupled to a gauge field whose dynamics was governed by the Chern-Simons term. The corresponding general coordinate invariant form for the Chern-Simons piece contains a correction that does not involve the  ${\cal{B}}$ field (\ref{gabbar1}). The ${\cal{B}}$- field now has a geometric role since it gets related to the spin connection (\ref{defb}) that is useful in understanding the geometry of FQHE.

Keeping the above observations in mind, let us now compare our results with that of \cite{SW}. This will be done in some details. 
 Apart from showing the connection of our approach with \cite{SW}, 
 it will also illustrate how some of the shortcomings or pitfalls there are avoided in the present context. The authors of \cite{SW}  
 obtained spatial diffeomorphism by applying the minimal coupling prescription to a theory of complex scalars leading to the action,   
\begin{equation}\label{free-L}
  S = \int dx^0 dx \sqrt{g}\left[\frac{i}{2} (\phi^{\dagger}\partial_{0}\phi-\psi\partial_{0}\phi^\dagger)
  - A_0\phi^{\dagger}\phi 
  - \frac{g^{ij}}{2m}(\partial_i\phi^{\dagger}-iA_i\phi^{\dagger})(\partial_j\phi+iA_j\phi)\right],
\end{equation}
which is invariant under infinitesimal transformations,
\begin{align}\label{3d-gci}
  x^i \to x^{i'} &= x^{i'}(x^i), \quad
  \phi(x^0,x) \to \phi(x^0,x') = \phi(x^0,x),\nonumber\\ 
  A_0(x^0,x)\to A_0'(x^0,x') &= A_0 (x^0,x), \quad
A_i(x^0,x) \to A_{i'} (x^0,x') = \frac{\partial x^i}{\partial x^{i'}}
    A_i(x^0,x)\nonumber\\  g_{ij}(x^0,x) \to g_{i'j'}(x^0,x') &= \frac{\partial x^i}{\partial x^{i'}}
    \frac{\partial x^j}{\partial x^{j'}} g_{ij}(x^0,x).
\end{align}
 when the fields transform as {\footnote{Note that, to make a comparison, we have set the gauge parameter in \cite{SW} to zero, since we consider only diffeomorphism symmetry.}},
\begin{align}\label{static-gci}
  \delta\phi &=- \xi^k\partial_k\phi, 
 \quad \delta A_0 = -  \xi^k\partial_k A_0,\nonumber\\
  \delta g_{ij} &= -\xi^k \partial_k g_{ij} - g_{ik}\partial_j \xi^k -
     g_{kj}\partial_i \xi^k, \quad
  \delta A_i =- \xi^k\partial_k A_i - A_k \partial_i\xi^k .
  \end{align}
 The action (\ref{free-L}) agrees with (\ref{diffaction3}) with the covariant derivative defined as (\ref{P11}). In the time independent case the transformations of basic fields given above becomes identical with that obtained here in (\ref{phic}, \ref{delA0curved}, \ref{delAicurved}, \ref{delBicurved}).
 
When the diffeomorphism parameter $\xi^i$ is time dependent the real difference comes up. The action (\ref{free-L}) is no longer invariant. Lacking a systematic algorithm the authors of \cite{SW} 
 found, ``by trial and error", that the invariance is restored by modifying the transformation of the gauge field as,
 \begin{align}
 \delta A_0&= -\xi^k \partial_k A_0- \dot{\xi}^kA_k\notag\\
 \delta A_i&= -\xi^k \partial_k A_i-A_k\partial_i\xi^k +mg_{ik}\dot{\xi}^k\label{Q}
 \end{align}
The other transformations in (\ref{static-gci}) are preserved. 
 
In our approach, on the other hand, we have a systematic algorithm. The transformations of all the fields are canonical in the sense that these are the standard ones found under general coordinate transformations. No ``trial and error" approach is needed. Thus, our transformations are given by (\ref{static-gci}) for $\phi, g_{ij}$ and $A_i$, while it is given by (\ref{Q}) for $A_0$. Also, one has to remember, that for time dependent transformations the variable $\Psi^k\neq0$ and the correct diffeomorphism invariant action is given by, 
\begin{eqnarray}
\tilde{S}=S+S_{\Psi} &=& \int dx^0 d^2x \sqrt{g}[ \frac{i}{2}\left(\phi^{*}{\bar{D}}_{0}\phi-\phi{\bar{D}}_0\phi^{*}\right) -g^{kl}\frac{1}{2m}\bar{D}_k\phi^{*}\bar{D}_l\phi]
\nonumber\\&+&\int dx^0 d^2x \sqrt{g} [\frac{i}{2}\Psi^k\left(\phi^{*}{\bar{D}}_{k}\phi
-\phi{\bar{D}}_k\phi^{*}\right)]
\label{diffaction12}
\end{eqnarray}
Expectedly, the new field $\Psi^k$ has an important role in discussing the complete spacetime transformation and is used in the construction of the basic elements of the Newton-Cartan geometry (section 4.8).

It is instructive to see how the invariance of $\tilde{S}$ is attained. Under the (canonical) set of transformations that we employ, the usual action $S$ changes as,
\begin{equation}
\delta S=-\frac{i}{2} \int dx^0 d^2x \sqrt{g}\dot{\xi}^k\phi^*(\overleftrightarrow{\bar{D}}_k)\phi\label{XX}
\end{equation}
To compute $\delta S_{\Psi}$ one has to specify the transformation property of $\Psi^k$. This is easily read-off from (\ref{delth1n}) by putting $\xi^0=0$ (vanishing temporal translation) and recalling that the boost parameter $v^i$ vanishes in the local frame, so that,
\begin{equation}
\delta \Psi^k=-\xi^i\partial_i\Psi^k+\Psi^m\partial_m \xi^k+\dot{\xi}^k
\end{equation}
We then find,
\begin{equation}
\delta S_{\Psi}=\frac{i}{2} \int dx^0 d^2x \sqrt{g}\dot{\xi}^k\phi^*(\overleftrightarrow{\bar{D}}_k)\phi
\end{equation}
which exactly cancels (\ref{XX}).

In the approach of \cite{SW} the usual (minimally coupled) action $S$ is retained. Then the change (\ref{XX}) is canceled by modifying the variation of $A_i$ by an additional (noncanonical) piece $mg_{ik}\dot{\xi}^k$ in (\ref{Q}).
To sum up, note that we do not demand any special transformation for the time dependent case. Identical transformation laws for the basic fields ensure the invariance of the action (\ref{diffaction12}). This is to be contrasted with \cite{SW} where the same action is retained but the transformation rules of the basic fields change in a non-canonical way {\footnote{These are given in (\ref{Q}) and correspond to equation (17) of \cite{SW}.}}. This is not surprising because the results of \cite{SW} are obtained in an adhoc manner. On the other hand our analysis does not distinguish between time dependent and time independent cases, both of which can be obtained in a holistic manner following our localization procedure.

Before finishing this comparison we would like to draw attention to a crucial point. This concerns the abstraction of flat limit. In our case we just replace the covariant derivatives by the ordinary ones and set the metric flat. The flat limit is smoothly recovered and does not depend on the time dependence (or independence) of $\xi^i$. 
A simple inspection of (\ref{diffaction12}) and (\ref{globalaction2}) confirms the above statement \footnote{Note that $\Psi^k$ vanishes when the covariant derivative is replaced by the ordinary derivative.}. In the approach of \cite{SW}, however, taking the flat limit is problematic due to the presence of the noncanonical piece $mg_{ik}\dot{\xi}^k$ in (\ref{Q}). Its flat limit would be $m\delta_{ik}\dot{\xi}^k$. Obviously such a term does not exist. The way out of this impasse was to include gauge transformations also so that,
\begin{equation}
\delta A_i=\partial_i\alpha-\xi^k \partial_k A_i-A_k\partial_i\xi^k+mg_{ik}\dot{\xi}^k\label{AA}
\end{equation}
The flat space Galilean transformations are then recovered with a specific choice \cite{SW},
\begin{equation}
\alpha=mv^i x^i,~~\xi^i=v^it\label{BB}
\end{equation}
so that,
\begin{equation}
\delta A_i=-tv^k\partial_k A_i\label{CC}
\end{equation}
yielding the expected transformation. Note that, to cancel the unwanted term in (\ref{AA}), it is essential to assume a particular relation between the gauge parameter and the boost parameter. This can hardly be motivated on fundamental premises. Such a shortcoming is avoided in our systematic approach. Of course we can also easily implement gauge transformations which is discussed in the next section.
\subsection{Gauge invariance for complex Schrodinger field theory in the presence of an external vector field}
The original action (\ref{action2})had a gauge invariance (\ref{gt}). The process of localisation, eventually leading to the diffeomorphism invariant action (\ref{diffaction12}) preserves this invariance.
An explicit demonstration of the gauge invariance of the action (\ref{diffaction12}) is straightforward. Let us first consider the structure of the derivatives appearing in (\ref{dbar}). Then under the gauge transformation (\ref{gt}) it is easy to show that these derivatives transform covariantly.
\begin{equation}
\bar{D}_{0}\phi\rightarrow (1+i\Lambda)\bar{D}_{0}\phi,~~~\bar{D}_{k}\phi\rightarrow (1+i\Lambda)\bar{D}_{k}\phi
\label{cog}
\end{equation}
Note that the new fields $(\cal{B})$  do not transform under the gauge transformation. Indeed if ${\cal{B}}$ changes under gauge transformation then the above covariant property is lost. The point is that the introduction of ${\cal{B}}$ was a consequence of the localization of spacetime symmetry. So ${\cal{B}}$ changes under general coordinate transformation but not under the gauge transformation. It may be recalled that the original gauge symmetry of the model is preserved by the process of localisation ( See for instance the discussion below {\ref{localaction}). 

Using the covariant property of the derivatives (\ref{cog}) it is easy to show that the action (\ref{diffaction12}) is invariant under the gauge transformation (\ref{gt}).
\subsection{Inclusion of the Chern-Simons term in the action}
Another landmark problem is the inclusion of the Chern-Simons (CS) term in the action \cite{HS, Son} because, as we have mentioned earlier, there is a certain confusion about the possibility of coupling the C-S action with curved 2- dim space \cite{HS, F}. The C-S action, due to its topological origin is known to be metric independent. So, the confusion has a surprise element. However, in our method the C-S action is smoothly included. We will see this in the following analysis.

 The CS action is given by 
\begin{equation}
S_{CS} = \int d^3x \frac{\kappa}{2}\epsilon^{\mu\nu\lambda}A_\mu\partial_\nu A_\lambda
\end{equation}
and can be coupled with both relativistic and non-relativistic models. It will be convenient to break the action in spatial and temporal parts,
\begin{equation}
S_{CS} = \int dx^0  \int d^2x_k \frac{\kappa}{2}\epsilon^{ij}\left(A_0\partial_i A_j-A_i\partial_0 A_j+A_i\partial_j A_0\right)
\label{globalactioncs} 
\end{equation} 

 It can be shown that (\ref{globalactioncs}) is invariant under the global Galilean transformation using the variations (\ref{delA}). Following the method to localize the Galilean transformation stated in the previous section, we can get the corresponding action invariant under the the local Galilean transformations as
\begin{align}
S = &\int dx^{\bar{0}} \int d^2x_a \frac{M}{\theta}
\frac{\kappa}{2}\epsilon^{ab}\left(A_{\bar{0}}\nabla_a A_b-A_a\nabla_{\bar{0}} A_b+A_a\nabla_b A_{\bar{0}}\right)
\label{localscaction} 
\end{align}
By our construction
this action (\ref{localscaction}) is invariant under (\ref{localgalilean}). This can also be checked explicitly.

 Now our algorithm given in section 3 allows us to construct the spatially diffeomorphic action as follows: 
\begin{eqnarray}
S &=& \int dx^0 d^2x \sqrt{g}\frac{\kappa}{2}\epsilon^{ab}{\Sigma_a}^k{\Sigma_b}^l\left[\left(A_0D_k A_l-A_kD_0 A_l+
A_kD_lA_0\right)\right.\nonumber\\ 
&+&\left.\Psi^m A_m D_k A_l+\Psi^m A_k\left(D_l A_m - D_m A_l\right)\right]
\label{curvedaction} 
\end{eqnarray}
Note that $\epsilon^{ab}$ is a tensor under local (orthogonal) transformations. Thus
\begin{equation}
{\Sigma_a}^k{\Sigma_b}^l\epsilon^{ab} ={\tilde{\epsilon}^{kl}} 
\end{equation}
where $\tilde{\epsilon}^{kl}$ is the Levi Civita tensor in the curved space. It is related to the numerical tensor 
$\epsilon^{kl}$ by,
\begin{equation}
\tilde{\epsilon}^{kl}=\frac{1}{\sqrt g}\epsilon^{kl}
\end{equation}

Then the CS action in curved space is obtained from the above equations as,
\begin{eqnarray}
S &=& \int dx^0 d^2x \frac{\kappa}{2}{{\epsilon}^{kl}}\left[\left(A_0D_k A_l-A_kD_0 A_l+
A_kD_lA_0\right)\right.\nonumber\\ 
&+&\left.\Psi^m A_m D_k A_l+\Psi^m A_k\left(D_l A_m - D_m A_l\right)\right]\label{curvedaction1}
\end{eqnarray}
Now the derivatives $D_{\mu}A_{\nu}$ are substituted from (\ref{covexpand}).
\begin{align}
S &=\int dx^0 d^2x \frac{\kappa}{2}{{\epsilon}^{kl}}\left[\left(A_0 (\partial_k A_l-\partial_l A_k+i{\cal{B}}_kA_l-i{\cal{B}}_lA_k)-A_k(\partial_0 A_l-\partial_l A_0+i {\cal{B}}_0 A_l-i{\cal{B}}_lA_0)\right. \right.\nonumber\\  & \left. \left.  +
A_k(\partial_l A_0-\partial_0 A_l+i{\cal{B}}_l A_0-i{\cal{B}}_0 A_l)\right) +
\Psi^m[ A_m (\partial_k A_l-\partial_l A_k+i{\cal{B}}_kA_l-i{\cal{B}}_lA_k)\right.\notag\\& \left.+ A_k(\partial_l A_m-\partial_m A_l+i{\cal{B}}_l A_m-i{\cal{B}}_mA_l)- A_k(\partial_m A_l-\partial_l A_m+i{\cal{B}}_m A_l-i{\cal{B}}_lA_m)\right]]\label{gabbar}
\end{align}
Exploiting the antisymmetric property of ${\epsilon}^{kl}$, (\ref{gabbar}) further reduces to,
\begin{align}
S &=\int dx^0 d^2x \frac{\kappa}{2}{{\epsilon}^{kl}}\left[2\left(A_0\partial_k A_l-A_k\partial_0 A_l+
A_k\partial_lA_0\right)\right.\nonumber\\ 
&+\left.2\Psi^m[ A_m \partial_k A_l+ A_k(\partial_l A_m-\partial_m A_l)\right]]\notag\\&=\int dx^0 d^2 x \kappa 
[\epsilon^{\mu\nu\lambda} A_{\mu}\partial_{\nu}A_{\lambda}+\Psi^m \epsilon^{kl}[ A_m \partial_k A_l+ A_k(\partial_l A_m-\partial_m A_l)]]
\label{gabbar1}
\end{align}
Note that the ${\cal{B}}$ field has dropped out from the above expression. Effectively, therefore, the Chern-Simons interaction receives a correction to its original form. At the first sight one may wonder about the correction term in view of the metric independence of the C-S action, But there is no surprise because we have implemented only diffeomorphism in the coordinate space. There is no space-time diffeomorphism. The correction term in (\ref{gabbar1}) rather owes its existence to the metric independence of the C-S action.

The correction term is infarct no hindrance to the general coordinate transformations. It may be shown that the  action (\ref{gabbar1}), under the general coordinate transformations (\ref{delAicurved}), (\ref{delA0curved}) and (\ref{varcurvedcov}), changes as
\begin{equation}
\delta S =\int dx^0 d^2x \kappa \partial_i\left[\xi^i\epsilon^{kl}\left(A_0\partial_kA_l - A_k\partial_0A_l + A_k\partial_lA_0\right)\right]
\end{equation}
The integrand is a total derivative and drops to zero when integrated over space. This proves that the action is invariant under the general coordinate transformations.

The Chern - Simons action has proved to be very useful in the study of fractional quantum Hall effect. In this context it may be noted that the Chern - Simons action is reported \cite{HS} to break the diffeomorphism symmetry. This has been a major obstacle in applying theories with Chern - Simons term in curved space. To recover the lost invariance it is essential to introduce correction fields. In our opinion these features are manifestations of the ad hoc prescription used to achieve non relativistic diffeomorphism invariance from a theory defined in flat space. Our approach on the other hand naturally leads to an appropriate Chern - Simons theory in curved space, without any adhoc assumptions.

\subsection{The question of gauge invariance in Chern-Simons interaction}
 Under the gauge transformation (\ref{gt}) the action (\ref{gabbar1}) can be shown to be invariant. The first piece is identically the Chern-Simons term whose gauge invariance is well known. The terms in the second parenthesis give a correction to the Chern-Simons action which will vary under the gauge transformation as,
\begin{align}
\delta {\cal{L}}&=2\Psi^m \epsilon^{kl}[(\partial_m\Lambda)(\partial_k A_l)+(\partial_k \Lambda)(\partial_l A_m-\partial_mA_l)\notag\\&=2\epsilon^{kl}[\partial_m(\Psi^m \Lambda \partial_k A_l)+\partial_k(\Psi^m \Lambda(\partial_l A_m-\partial_m A_l))\notag\\&-\Lambda[(\partial_m\Psi^m)(\partial_k A_l)+(\partial_k \Psi^m)(\partial_l A_m-\partial_m A_l)]]
\end{align}
The second term proportional to $\Lambda$ vanishes identically. Thus $\delta{\cal{L}}$ is a pure boundary so that the action (\ref{gabbar1}) remains invariant.

Note that $\Psi^m$ which appears in the above example is actually related to the Newton-Cartan data as will be seen below \cite{BMM2}. 
\subsection{The applications of Chern Simons theory in FQHE}
We will now demonstrate that  there are no particular difficulties in using the Chern Simons theory in the non relativistic diffeomorphism invariant space. Recently such theories were used by casting the flat space version of C-S gauge theories to a sheared space where the deformation is independent of time \cite{F}. They used a scalar field theory interacting with the C-S field. Much efforts were needed to explain the coupling of the scalar field with curved space time.  We will find that in our formalism the scalar field is most naturally coupled to the spin connection. The form of the covariant derivative agrees with that of \cite{F}.

 In two space dimensions there is only one component of angular momentum that generates rotation in space. The rotation group is therefore abelian. This is the simplest way one can understand why the excitations can have arbitrary (fractional) spin. The spin statistics connection then shows that the statistics is also arbitrary. This fractional spin can be attributed to a nonrelativistic electron by attaching flux - charge composites to the particle which are vortices of the C-S theory \cite{BMo}. Thereby one can view the particle as a composite fermion or a composite boson 
 \cite{F}. In flat space the particle action is given by the Lagrangian   
\begin{equation}
S = \int dx^0  \int d^2x_k  \left[ \frac{i}{2}\left( \phi^{*}\Delta_{0}\phi-\phi \Delta_0\phi^{*}\right) -\frac{1}{2m}\Delta_k\phi^{*}\Delta_k\phi\right] + \int dx^0 d^2 x \kappa 
\left[\epsilon^{\mu\nu\lambda} a_{\mu}\partial_{\nu}a_{\lambda}\right]
\label{globalaction2n} 
\end{equation}
where
\begin{eqnarray}
\Delta_{\mu}\phi = \partial_\mu\phi + iA_\mu \phi : A_\mu = A_
\mu^{ext}+a_\mu 
\end{eqnarray}
The gauge field $A_{\mu}$ contains the external field which produces the constant magnetic field in the $z$ direction and the topological Chern Simons field ($a_\mu$) that gives rise to the flux tube attachment with the electron. Now the above theory is invariant under global Galilean transformations (\ref{globalgalilean}).
Following our algorithm it is an easy exercise to couple (\ref{globalaction2n}) with curved space - time. As a first step, the partial derivatives should be replaced by 
covariant derivatives. For convenience we will rename
\begin{eqnarray}
{\Sigma_0}^{0} =\theta \hskip .2cm \rm{and} \hskip .2cm {\Sigma_0}^{k}=\Psi^k\label {newdefsigma}
\end{eqnarray}
This nomenclature will be useful in the subsequent analysis. The covariant derivatives are now given by
\begin{equation}
\nabla_{\alpha}\phi={\Sigma_\alpha}^{\mu}D_{\mu}\phi\label{fcovdev}
\end{equation}
where $D_{\mu}\phi$ is expressed as
\begin{equation}
D_{\mu}\phi = \partial_{\mu}\phi + i{\cal{B}}_{\mu}\label{covdev}
\end{equation}
The measure of the volume of integration transforms to \cite{BMM1}
\begin{equation}
\int d^3x \longrightarrow \int d^3x \frac{\det{\Lambda_k^{{}{a}}}}{\Sigma_0^{{}{0}}}\label{measuren}
\end{equation}
$B_{\mu}$ is a new gauge field introduced in the localisation process and $\Lambda_\mu^{{}\alpha}$ is the inverse of ${\Sigma_\alpha}^{\mu}$. Substituting 
these changes the action (\ref{globalaction2n}) can be reinterpreted as a theory in the curved space. Note that for this interpretation to be valid with the canonical transformation of the fields, the spatial deformation $\xi^i$
must be time independent, so that $\Sigma_0^{{}0} =1$ and $\Sigma_0^{{}k} =0$ \cite{BMM3}.

 The action of the C-S coupled non relativistic electron theory in curved space can be written immediately from the above algorithm. It is given by, 
\begin{equation}
S = \int dx^0  \int d^2x_k \sqrt{g} \left[ \frac{i}{2}\left( \phi^{*}\bar{D}_{0}\phi-\phi \bar{D}_0\phi^{*}\right) -\frac{1}{2m}g^{kl}\bar{D}_k\phi^{*}\bar{D}_l\phi\right] + \int dx^0 d^2 x \kappa 
\left[\epsilon^{\mu\nu\lambda} a_{\mu}\partial_{\nu}a_{\lambda}\right]
\label{localation} 
\end{equation}
where
\begin{eqnarray}
\bar{D}_{\mu}\phi = D_{\mu}\phi+iA_\mu\phi=\partial_\mu\phi +iA_\mu + i{\cal {B}}_\mu\phi
\label{dbarn}
\end{eqnarray}
and
$g$ is the determinant of $g_{ij}$ with
\begin{equation}
g_{ij} ={ \Lambda_i}^{a}{ \Lambda_j}^{b}\delta{ab}
\end{equation}
where $a,b$ are local coordinate labels on the tangent space.
The theory (\ref{localation}) can easily be interpreted as a theory in curved space with metric $g_{ij}$. From the transformation rules of ${\Sigma_a}^k$ it can be easily shown that $g_{ij}$ transforms as a second rank tensor under spatial diffeomorphism.

 In this geometric backdrop $B_\mu$ is related to the spin connection. Recalling (\ref{gaugefields}) and the fact that the diffeomorphism parameter is independent of time, we write,
\begin{equation}
{\cal{B}}_\mu\phi = \frac{1}{2}{\cal{B}}_\mu^{ab}\omega_{ab}\phi\label{connection}
\end{equation}
where $\omega_{ab}$ is the spin part of the rotation generator. Since we are working in two space dimensions the right hand side of (\ref{connection}) is equal to ${\cal{B}}_\mu^{12}\omega_{12}$. For the vortices of (\ref{globalaction2n})the angular momentum operator $\omega_{12} = K$, where $K$ is the topological spin \cite{BMo}. The single component spin connecton thus may be written without the local indices and,
\begin{equation}
{\cal{B}}_{\mu}\phi=K\omega_{\mu}^{12}\phi\label{defb}
\end{equation}
The structure of the covariant derivative ($ \bar D_\mu$) found in \ref{dbarn} with $B_{\mu}$ defined as (\ref{defb}) is identical to the form suggested in \cite{F}. This form of the covariant derivative reproduces the Hall viscosity and Wen Zee shift, as shown in \cite{F}. Though our results agree with that of \cite{F} there is a significant difference. In \cite{F}, there is a long discussion followed by elaborate deductions to justify how the scalar field couples with the background geometry. In our theory it follows simply from geometry.
\subsection{Hydrodynamical form of Schroedinger theory}
An interesting and novel application of our approach is revealed in the context of fluid dynamics. 
We start with the Schr\"odinger lagrangian with a potential term
\begin{equation}
S = \int dt  \int d^3x  \left[ \frac{i}{2}\left( \phi^{*}\partial_{t}\phi-\phi\partial_t\phi^{*}\right) -\frac{1}{2m}\partial_k\phi^{*}\partial_k\phi- V\left(\phi^*\phi\right)\right].
\label{globalactiong} 
\end{equation} This is just the the action (\ref{globalaction}), augmented by a self interaction. If we express the complex field $\phi$ in polar variables, 
\begin{equation}
\label{a}
\phi = \sqrt{\rho} e^{i \alpha}
\end{equation}
then the action (\ref{globalactiong}) takes the form,
\begin{equation}
\label{b2}
S=\int dt \int d^{3}x\left[-\rho\dot{\alpha}-\left(\dfrac{\rho}{2m}(\partial_{k}\alpha)^{2}+\dfrac{1}{8m\rho}(\partial_{k}\rho)^{2}\right)- \bar{V}\left(\rho\right)\right].
\end{equation}
where,
\begin{equation}
\label{d1}
{\bar{V}}(\rho)= V(\rho) + \dfrac{1}{8m\rho}(\partial_{k}\rho)^{2}
\end{equation}
This is usually referred as the hydrodynamical version of Schroedinger theory. The fluid variables are the density $\rho$ (which is now the analogue of the matter density $\phi^*\phi$) and the velocity $v_{k}$ that is related to $\alpha$ by,
\begin{equation}
\label{c1}
v_{k}=\partial_{k}\alpha.
\end{equation}
Incidentally, the above relation is the Clebsch decomposition of the velocity for an irrotational fluid $(\vec{\nabla}\times\vec{v}=0)$. Since the lagrangian (\ref{b2}) is already in the first order form, $(\rho,\alpha)$ is the canonical pair in the hamiltonian formulation. The hamiltonian for the fluid is also easily read-off from (\ref{b2}).It involves a corrected potential term $\bar{V}$
even if the original Schr\"oedinger theory (\ref{globalactiong}) did not have any such term (i.e. $V(\rho) =0)$. Furthermore, since entropy is not involved, the hydrodynamics emenating from the Schroedinger theory is an example of irrotational and isentropic fluid.

Let us next consider the symmetries of the model (\ref{b2}). It is obvious that it possesses a Galilean symmetry. This is easily manifested by considering the symmetry in the original model (\ref{globalactiong}). Using the variable change (\ref{a}) the form variation of the fluid variables comes out as,
\begin{eqnarray}
\label{e}
\delta_{0}\rho &=& -\xi^{\mu}\partial_{\mu}\rho   \\ \nonumber
\delta_{0}\alpha &=& -\xi^{\mu}\partial_{\mu}\alpha -mV^{i}x_{i}.
\end{eqnarray}
The inhomogeneous contribution in $\delta_{0}\alpha$ is the familiar Galileo $1-$cocycle corresponding to the boost algebra.

Apart from the Galilean symmetry a nontrivial symmetry connected to a field dependent diffeomorphism transformation was reported in \cite{BJ}.
However, its interpretation or geometric meaning was unclear. We, however, can easily provide the diffeomorphic form of (\ref{b2})in space from the systematic approach of obtaining nonrelativistic diffeomorphism in space, which has been developed above. 

As we know, the algorithm to incorporate spatial NRDI in (\ref{globalactiong}) contains two steps. First, we get a local galilean invariant theory in flat space  by replacing the partial derivatives by appropriate covariant derivatives and, furthermore, suitably altering the measure. This yields, from \ref{ssl}),the theory
\begin{align}
S = \int dx^{\bar{0}} \int d^2x_a \frac{M}{\theta}&\left[\frac{i}{2}\left(\phi^{*}\nabla_{\bar{0}}\phi-\phi \nabla_{\bar{0}}\phi^{*}\right)-\frac{1}{2m}\nabla_a\phi^{*}\nabla_a\phi - V\left(\phi^*\phi\right)
\right]
\label{localscintactiong} 
\end{align}
where,
$\nabla_{\bar{0}}\phi$ and $\nabla_a\phi$ are given by 
(\ref{finalcov}) and (\ref{nab}), respectively. Equation (\ref{localscintaction}) is an action which is invariant in flat space under the transformations (\ref{delphi}), (\ref{delth1}) and (\ref{delth1n}). 

 It has been discussed that the transformation equations (\ref{delth1})  and (\ref{delth1n}) point towards a geometrical interpretation. Instead of viewing the fields $\Sigma_a^{{}k}$ defined in flat space, we can consider them to be the vielbeins connecting the coordinate basis with the local orthogonal basis in the patch containing a spatial point.  Thus the locally galilean symmetric theory (\ref{localscintactiong}) can again be interpreted as having a symmetry under  general coordinate invariance in space. The approriate extension of (\ref{globalactiong}) to a theory with non relativistic invariance in curved space is then given by the action
\begin{eqnarray}
S &=& \int dx^0 d^2x \sqrt{g}[ \frac{i}{2}\left(\phi^{*}D_{0}\phi-\phi D_0\phi^{*}\right) -g^{kl}\frac{1}{2m}D_k\phi^{*}D_l\phi
\nonumber\\&+& \frac{i}{2}\Psi^k\left(\phi^{*}D_{k}\phi
-\phi D_k\phi^{*}\right)- V\left(\phi^*\phi\right)]
\label{diffaction12g}
\end{eqnarray}
where $D_k\phi$ and $D_0\phi$ are defined in (\ref{dkphi})
 
Now sustituting $\phi = \sqrt{\rho} e^{i\alpha}$ in (\ref{diffaction12g}) and recalling that for discussing spatial NRDI, $\theta=1$, we get after some calculations, 
\begin{align}
S = \int d^3x \sqrt{-g}&\left[\frac{i}{2}\left( \rho \left(\partial_0\alpha + {\cal{B}}_0\right)+ \Psi^k\rho\left(\partial_k\alpha  +{\cal{B}}_k\right) \right)-\frac{1}{2m}g^{kl}\left(\partial_k\alpha + {\cal{B}}_k\right)\left(\partial_l\alpha + {\cal{B}}_l\right) - \bar{V}\left(\rho\right)
\right]
\label{calscintaction} 
\end{align}
which is the NRDI fluid model corresponding to (\ref{b2}). It can be easily shown from (\ref{phi2}) and (\ref{delkphin}) that both $\rho$ and $\alpha$ transform as scalar with respect to non relativistic spatial diffeomorhism symmetry. 

Note that the passage to the flat limit of (\ref{calscintaction}) is trivial. One just has to substitute $g_{ij}$ by $\delta_{ij}$ and put the new fields $B_0$, $B_k$
and $\Psi_k$ to be zero. Remarkably, this prescription of going to the flat limit is universal in our method of obtention of NRDI in space. This is a welcome feature in view of the confusion prevalent in the literature.
\subsection{Space-time transformations} 
The obtention of nonrelativistic diffeomorphism invariance algorithm developed in \cite{BMM1, BMM3} is completely general. In the above we have considered a subset where the time translation parameter $\epsilon = 0$. Naturally, a question arises as to how our algorithm corresponds to the Galileo - Newton space time, the metric formulation of which was done by E. Cartan \cite{Cartan-1923, Cartan-1924} and perfected over the years \cite{Havas} - \cite{MALA} by many stalwarts. We will show that the Newton Cartan geometry in the most general form (torsion or without torsion) is directly obtained by our method. The story does not end here. In course of this discussion we will see that the perspective of the non relativistic diffeomorphism developed in our formalism can be used to obtain another nonrelativistic geometry -- the much discussed Horava - Lifshitz \cite{H} geometry. Interestingly but not accidentally, the flat limit symmetries of the H-L geometry comes out to be the Galilean algebra without boost, unlike the Newton - Cartan which has full Galilean symmetry.

 In the present section we will consider the geometric interpretation of the localisation of Galilean symmetry in its full glory. We define a four dimensional manifold, the `coordinates' of which are denoted by $x_0, x_1, x_2, x_3$. Set up local and global frames. Redefining the new fields introduced in the previous section as in (\ref{newdefsigma})
and putting ${\Sigma_a}^0=0$ we can construct the $4\times 4$ invertible matrix,
\begin{equation}
{\Sigma_\alpha}^{\mu}=\begin{pmatrix}
\theta & \theta\Psi^k \\
0 & {\Sigma_a}^k
\end{pmatrix}
\label{Smatrix1}
\end{equation}
where ${\Sigma_a}^k$ has already been introduced in (\ref{nab}). The inverse matrix ${\Lambda_\mu}^\alpha$ satisfies,
\begin{equation}
{\Sigma_\alpha}^{\mu}\Lambda_{\mu}{}^{\beta}=\delta^{\beta}_{\alpha},~~~
{\Sigma_\alpha}^{\mu}\Lambda_{\nu}{}^{\alpha}=\delta^{\mu}_{\nu}
\label{sila}
\end{equation}
The spatial part $\Lambda_k{}^a$ is the inverse of $\Sigma_a{}^k$ as may be seen in (\ref{sl}). Note that we are denoting the local coordinates by the initial Greek letters i.e. $\alpha,\beta $ etc. whereas the global coordinates are denoted by letters from the middle of the Greek alphabet i.e.  $\mu,\nu $ etc. From the definitions of ${\Sigma_\alpha}^{\mu}$ and $\Lambda_{\mu}{}^{\alpha}$ and the transformations of the various fields involved we can obtain the corresponding transformation laws. Thus,
\begin{align}
\delta_0 {\Sigma_0}^{k} &=\epsilon{\dot{\Sigma}_0}^{k}+ {\Sigma_0}^{i}\partial_{i}(\eta^{k}-tv^{k}) - \eta^{i}\partial_{i}{\Sigma_0}^{k}+t v^{i} \partial_{i} {\Sigma_0}^{k}
+\partial_0\left(\eta^k - tv^k\right)\theta + v^b{\Sigma_b}^{k}\nonumber\\
\delta_0 {\Sigma_a}^{k} &=\epsilon{\dot{\Sigma}_a}^{k}+ {\Sigma_a}^{i}\partial_{i}\eta^{k}-t{\Sigma_a}^{i}\partial_{i}v^{k} -
\lambda^b_{a}{\Sigma_b}^{k} - \eta^{i}\partial_{i}{\Sigma_a}^{k}+t v^{i} \partial_{i} {\Sigma_a}^{k}
\label{delth2}
\end{align}
Using the definition of $\xi^i$ from (\ref{localgalilean}) the above equations can be simplified to,
\begin{align}
\delta_0 {\Sigma_0}^{k} &= -\xi^\nu {\partial_\nu{\Sigma}_0}^{k}+ {\Sigma_0}^{\nu}\partial_{\nu}\xi^{k} - v^b{\Sigma_b}^{k}\nonumber\\
\delta_0 {\Sigma_a}^{k} &= -\xi^\nu {\partial_\nu{\Sigma}_a}^{k}+ {\Sigma_a}^{\nu}\partial_{\nu}\xi^{k} -\lambda_a{}^b{\Sigma_b}^{k}
\label{delth3}
\end{align}
Similarly we can work out,
\begin{align}
\delta_0 {\Lambda_0}^{a} &= -\xi^\nu {\partial_\nu{\Lambda}_0}^{a}- {\Lambda_\nu}^{a}\partial_{0}\xi^{\nu} - v^a{\Lambda_0}^{0}\nonumber\\
\delta_0 {\Lambda_k}^{a} &= -\xi^\nu {\partial_\nu{\Lambda}_a}^{k}- {\Lambda_\nu}^a\partial_{k}\xi^{\nu} -{\lambda_c}^a{\Lambda_k}^{c}
\label{delth4}
\end{align}

The matrix (\ref{Smatrix1}) has been derived totally from a nonrelativistic perspective. Their elements transform from flat tangent space to global space time manifold in the neighborhood of a point on the manifold. In the flat space the spacetime symmetries are the local Galilean symmetries. In the following we will first obtain the Newton - Cartan geometry by appropriately constructing two degenerate metrics in correspondence with Cartan's prescription. This leads to the most general Newton - Cartan geometry with torsion. If we impose torsion free condition by symmetrizing the connection then the standard N-C geometry emerges.

 Interestingly, if we impose a nondegenerate metric structure in space time we get a metric theory  with NRDI. Since we have introduced a single nondegenerate metric the full galilean symmetry can not hold in the tangent space. Specifically, the Galilean boost ceases to be a symmetry. Remarkably, the geometric structure of the resulting spacetime is identical with the projectable version of the Horava - Lifshitz geometry. The nonrelativistic origin of the Horava - Lifshitz geometry and limitations of its symmetries are revealed with hitherto unachieved clarity. The present section will demonstrate the validity of these assertions.
 
\subsection{ Newton-Cartan Spacetime }

Before beginning the actual construction it will be useful to review the salient features of the Newton-Cartan spacetime \cite{Daut}. It is a four dimensional manifold ${\cal{M}}$ endowed with two degenerate metrices of rank 3 and rank 1 respectively. It is convenient to take a degenerate spatial metric $h^{\mu\nu}$ of rank 3 and a degenerate temporal vielbein $\tau_{\mu}$ of rank 1. A connection $\Gamma^{\mu}_{\nu\lambda}$ is assumed with respect to which the following metricity conditions hold,
\begin{equation}
\nabla_\mu h^{\mu\nu} = 0,~~~~\nabla_\mu \tau_{\nu} = 0.
\label{metricity}
\end{equation}
To get an explicit form of the connection $\Gamma^{\mu}_{\nu\lambda}$ we also require to introduce the covariant quantities $h_{\mu\nu}$ and the contravariant ones $\tau^\mu$. These are defined by the following properties
\begin{align}
 h^{\mu\nu}h_{\nu\rho} &= \delta^{\mu}_{\rho}-\tau^{\mu}\tau_{\rho},~~~~\tau^{\mu}\tau_{\mu} = 1,\nonumber\\
 ~~~h^{\mu\nu}\tau_{\nu} &= 0,~~~~ h_{\mu\nu}\tau^{\nu} = 0.
\label{inverse}
\end{align}
Using these the connection can be written as,
\begin{align}
{\Gamma^\sigma}_{\mu\nu} & = \tau^{\sigma}\partial_{(\mu}\tau_{\nu)} +
\frac{1}{2}h^{\sigma\rho} \Bigl(\partial_{\nu}h_{\rho\mu} + \partial_{\mu}h_{\rho\nu} - \partial_{\rho}h_{\mu\nu}\Bigr)
\label{ncm}
\end{align}
But the connection $\Gamma_{\mu\nu}^{\rho}$ is not uniquely determined by the metric compatibility conditions (\ref{metricity}). These conditions are also preserved under the shift,
\begin{equation}
\Gamma^{\rho}_{\mu\nu} \rightarrow \Gamma^{\rho}_{\mu\nu} + h^{\rho\lambda}K_{\lambda(\mu}\tau_{\nu)}
\label{connectionshift}
\end{equation}
where $K_{\mu\nu}$ is an arbitrary two-form \cite{Daut}. Now it is possible to write the most general form of the metric compatible symmetric connection using this arbitrary two-form and (\ref{ncm}) as \cite{Daut, B},
\begin{align}
{\Gamma^\sigma}_{\mu\nu} & = \tau^{\sigma}\partial_{(\mu}\tau_{\nu)} +
\frac{1}{2}h^{\sigma\rho} \Bigl(\partial_{\nu}h_{\rho\mu} + \partial_{\mu}h_{\rho\nu} - \partial_{\rho}h_{\mu\nu}\Bigr)+ h^{\sigma\lambda}K_{\lambda(\mu}\tau_{\nu)}\,.
\label{covariantconnection}
\end{align}
Our next task is to show that the 4-dimensional spacetime manifold endowed with the matrix $\Sigma_{\alpha}{}^{\mu}$ and its inverse $\Lambda_{\mu}{}^{\beta}$ has the features of the Newton-Cartan geometry provided all the (extended) Galilean symmetry elements are retained in the tangent space .
 With this point of view we write down a degenerate spatial metric $h^{\mu\nu}$ of rank 3 and a degenerate temporal vielbein $\tau_{\mu}$ of rank 1 in the following way
 \begin{equation}
h^{\mu\nu}={\Sigma_a}^{\mu}{\Sigma_a}^{\nu}
\label{spm}
\end{equation}
and
\begin{equation}
\tau_{\mu}={\Lambda_\mu}^{0}
\label{tem}
\end{equation}
From the form variations of the basic fields (see (\ref{delth1},\ref{delth2}, \ref{delLamb})) we get,
\begin{equation}
\delta_0 h^{\mu\nu} =-\xi^{\rho} \partial_{\rho} h^{\mu\nu}+h^{\rho\nu}\partial_{\rho}\xi^{\mu}+h^{\rho\sigma}\partial_{\sigma}\xi^{\mu}
\label{diffs} 
\end{equation}
Similar results can be obtained for $\delta_0 \tau_{\mu}$. 
\begin{equation}
\delta_0 \tau_{\mu}=\delta_0 \Lambda_{\mu}{}^0=\delta_0 \Lambda_0{}^0=-\Lambda_0{}^0\partial_0 \xi^0+\xi^0\partial_0 \Lambda_0{}^0
\end{equation}
Using these relations it is easy to show that they have correct tensorial properties,
\begin{equation}
 h^{\mu\nu}\left(x^{\prime}\right) = \frac{\partial x^{\prime\mu}}{\partial x^{\rho}}\frac{\partial x^{\prime\nu}}{\partial x^{\sigma}}h^{\rho\sigma}\left(x\right)
\label{diffh} 
\end{equation}
and
\begin{equation}
 \tau_{\mu}\left(x^{\prime}\right) = \frac{\partial x^{\rho}}{\partial x^{\prime\mu}}\tau_{\rho}\left(x\right)
\label{difft} 
\end{equation}
The quantities ${\Sigma_\alpha}^{\mu}$ and $\Lambda_{\mu}{}^{\alpha}$ can be considered as direct and inverse vielbein and the fields $B_\mu^{\alpha\beta}$ (\ref{gaugefields}) may be interpreted as the spin connection. The affine connection $\Gamma_{\nu\mu}^{\rho}$ is introduced through the vielbein postulate \cite{Daut}, 
\begin{equation}
\nabla_\mu{\Lambda^\alpha}_{\nu} = \partial_{\mu}{\Lambda^\alpha}_{\nu} - \Gamma_{\nu\mu}^{\rho}{\Lambda^\alpha}_{\rho}
+B^{\alpha}{}_{\mu\beta}{\Lambda^\beta}_{\nu} =0\,. 
 \label{P}
\end{equation}
For $\alpha=0$ from (\ref{P}) we can get,
\begin{equation}
\nabla_{\mu} \Lambda_{\nu}{}^0= \partial_{\mu}\Lambda_{\nu}{}^0 - \Gamma_{\nu\mu}^{\rho}\Lambda_{\rho}{}^0+B^{0}{}_{\mu\beta}\Lambda_{\nu}{}^{\beta} =0\,. 
 \label{P1}
\end{equation}
Since ${B^0}_{\mu\beta}$ vanishes for a Galilean transformation, (\ref{P1}) implies,
\begin{equation}
 \partial_{\mu}\Lambda_{\nu}{}^0 - \Gamma_{\nu\mu}^{\rho}\Lambda_{\rho}{}^0=0
\end{equation}
Now using (\ref{tem}),we find,
\begin{equation} 
 \nabla_{\mu}\tau_{\nu}=0
\label{tau}
\end{equation}
This proves the metricity condition (\ref{metricity}) for $\tau_{\mu}$.
 
The proof of the metricity condition for $h^{\mu\nu}$ is a little bit involved. From (\ref{P})using (\ref{sila}) it can be shown that,
\begin{equation}
\partial_{\mu}\Sigma_{\delta}{}^\sigma-B_{\mu}{}^{\beta}{}_{\delta}\Sigma_{\beta}{}^\sigma =-\Gamma_{\nu\mu}^{\sigma}\Sigma_{\delta}{}^\nu
\label{P2}
\end{equation}
Considering $\delta=a$ we get{\footnote{As $B_\mu{}^{0a}$ = 0},
\begin{equation}
\partial_{\mu}\Sigma_{a}{}^\sigma-B_{\mu}{}^{b}{}_{a}\Sigma_{b}{}^\sigma =-\Gamma_{\nu\mu}^{\sigma}\Sigma_{a}{}^\nu
\label{Ploc}
\end{equation}
Multiplying (\ref{Ploc}) by $\Sigma_{a}{}^{\rho}$ yields,
\begin{equation}
\Sigma_{a}{}^{\rho}\partial_{\mu}\Sigma_{a}{}^\sigma-B_{\mu}{}^{b}{}_{a}\Sigma_{a}{}^{\rho}\Sigma_{b}{}^\sigma =-\Gamma_{\nu\mu}^{\sigma}\Sigma_{a}{}^{\rho}\Sigma_{a}{}^\nu
\label{Plocmu}
\end{equation}
Then we interchange the indices $\rho, \sigma$,
\begin{equation}
\Sigma_{a}{}^{\sigma}\partial_{\mu}\Sigma_{a}{}^\rho-B_{\mu}{}^{b}{}_{a}\Sigma_{a}{}^{\sigma}\Sigma_{b}{}^\rho =-\Gamma_{\nu\mu}^{\rho}\Sigma_{a}{}^{\sigma}\Sigma_{a}{}^\nu
\label{Plocmu11}
\end{equation}
Adding (\ref{Plocmu}, \ref{Plocmu11}) and using the antisymmetric property of $B_{\mu}{}^{a b}$ we obtain,
\begin{equation}
\partial_{\mu}h^{\rho\sigma}+\Gamma_{\nu\mu}^{\rho}h^{\nu\sigma}+\Gamma_{\nu\mu}^{\sigma}h^{\nu\rho}=0
\end{equation}
where $h^{\mu\nu}$ is defined in (\ref{spm}). That means,
\begin{equation}
\nabla_{\mu}h^{\rho\sigma}=0
\label{spmetricity}
\end{equation}
Thus we can conclude that our constructions of $h^{\mu\nu}$ (\ref{spm}), $\tau_\mu (\ref{tem})$ and $\Gamma_{\nu\mu}^{\rho} (\ref{P})$ satisfy the metric compatibility conditions (\ref{metricity}).

Our next task is to entirely express the connection in terms of the metric. For this we require the covariant metric $h_{\mu\nu}$ and contravariant ${\tau^\mu}$. Let
\begin{equation}
h_{\nu\rho}=\Lambda_{\nu}{}^{a} \Lambda_{\rho}{}^{a}
\label{spm2}
\end{equation}
and
\begin{equation}
\tau^{\rho}={\Sigma_0}^{\rho}.
\label{tm2}
\end{equation}
Using (\ref{spm}) and (\ref{tem}) we immediately get,
\begin{align}
h^{\mu\nu}\tau_\nu &={\Sigma_a}^{\mu}{\Sigma_a}^{\nu} \Lambda_{\nu}{}^{0}\notag\\
&={\Sigma_a}^{\mu}\delta^0_a\notag\\ &=0
\end{align}
Also the identifications (\ref{tm2})and (\ref{tem}) show that 
$$\tau^\mu\tau_\mu = 1.$$From the definitions (\ref{spm2}) and (\ref{tm2}) we find
\begin{align}
h_{\mu\nu}\tau^\nu &=  {\Lambda_{\mu}}^a {\Lambda_\nu}^a {\Sigma_0}^\nu\notag\\ &=  {\Lambda_{\mu}}^a\delta_0^a\notag\\ &=0
\end{align}
Finally we can easily verify that $$h^{\mu\lambda}h_{\lambda\nu} = \delta^\mu_\nu - \tau^\mu\tau_\nu.$$
This completes the realization of the Newton-Cartan algebra given in (\ref{inverse}).

Finally, it will be shown that the connection $\Gamma_{\nu\mu}^{\rho}$ defined in (\ref{P}) can be put in the general form (\ref{covariantconnection}) as required in the Newton-Cartan construction. We can write from (\ref{P}), 
\begin{align}
\Gamma_{\nu\mu}^{\rho} &= \partial_{\mu}{\Lambda_{\nu}}^\alpha {\Sigma_\alpha}^{\rho}
+B^{\alpha}{}_{\mu\beta}{\Lambda_{\nu}}^\beta{\Sigma_\alpha}^{\rho}\notag\\
&=\frac{1}{2}[\partial_{\mu}{\Lambda_{\nu}}^0 {\Sigma_0}^{\rho}+\partial_{\nu}{\Lambda_{\mu}}^a {\Sigma_a}^{\rho}+ B^{\alpha}{}_{\mu\beta}{\Lambda_{\nu}}^\beta{\Sigma_\alpha}^{\rho}+B^{\alpha}{}_{\nu\beta}{\Lambda_{\mu}}^\beta{\Sigma_\alpha}^{\rho}]\notag\\ &=\frac{1}{2}[\partial_{\mu}{\Lambda_{\nu}}^{0} {\Sigma_0}^{\rho}+\partial_{\nu}{\Lambda_{\mu}}^0 {\Sigma_0}^{\rho}+\partial_{\mu}{\Lambda_{\nu}}^a {\Sigma_a}^{\rho}+\partial_{\nu}{\Lambda_{\mu}}^a {\Sigma_a}^{\rho}\notag\\ &+B^{a}{}_{\mu 0}{\Lambda_{\nu}}^{0}{\Sigma_\alpha}^{\rho}+B^{a}{}_{\nu 0}{\Lambda_{\mu}}^0{\Sigma_a}^{\rho}+B^{a}{}_{\mu b}{\Lambda_{\nu}}^b{\Sigma_\alpha}^{\rho}+B^{a}{}_{\nu b}{\Lambda_{\mu}}^b{\Sigma_a}^{\rho}]\notag\\
 \label{PP}
\end{align}

As, ${\Sigma_a}^{\rho}=h^{\rho\sigma}{\Lambda_{\sigma}}^a$, the above expression will take the form as,
\begin{align}
\Gamma_{\nu\mu}^{\rho}&=\frac{1}{2}[(\partial_{\nu}\tau_{\mu})\tau^{\rho}+(\partial_{\mu}\tau_{\nu})\tau^{\rho}
+(\partial_{\mu}{\Lambda_{\nu}}^a)h^{\rho\sigma} {\Lambda_\sigma}^{a}+(\partial_{\nu}{\Lambda_{\mu}}^a) h^{\rho\sigma}{\Lambda_\sigma}^{a}]\notag\\&+B^{a}{}_{0\mu}{\Lambda_{\nu}}^{0}{\Sigma_\alpha}^{\rho}+B^{a}{}_{0\nu}{\Lambda_{\mu}}^0{\Sigma_a}^{\rho}+B^{a}{}_{\mu b}{\Lambda_{\nu}}^b{\Sigma_\alpha}^{\rho}+B^{a}{}_{\nu b}{\Lambda_{\mu}}^b{\Sigma_a}^{\rho}\notag\\ &=\tau^{\rho}\partial_{(\mu}\tau_{\nu)}+
\frac{1}{2}h^{\rho\sigma}[\partial_{\mu}({\Lambda_{\nu}}^a{\Lambda_{\sigma}}^a)-{\Lambda_{\nu}{}^a}\partial_{\mu}{\Lambda_{\sigma}}^a]+\frac{1}{2}h^{\rho\sigma}[
\partial_{\nu}({\Lambda_{\mu}}^a{\Lambda_{\sigma}}^a)-{\Lambda_{\mu}{}^a}\partial_{\nu}{\Lambda_{\sigma}}^a]\notag\\& +B^{a}{}_{0\mu}{\Lambda_{\nu}}^{0}{\Sigma_\alpha}^{\rho}+B^{a}{}_{0\nu}{\Lambda_{\mu}}^0{\Sigma_a}^{\rho}+B^{a}{}_{\mu b}{\Lambda_{\nu}}^b{\Sigma_\alpha}^{\rho}+B^{a}{}_{\nu b}{\Lambda_{\mu}}^b{\Sigma_a}^{\rho}\notag\\&=\tau^{\rho}\partial_{(\mu}\tau_{\nu)}+
\frac{1}{2}h^{\rho\sigma}[\partial_{\mu} h_{\sigma\nu}-{\Lambda_{\nu}{}^a}\partial_{\mu}{\Lambda_{\sigma}}^a]+\frac{1}{2}h^{\rho\sigma}[
\partial_{\nu} h_{\sigma\mu}-{\Lambda_{\mu}{}^a}\partial_{\nu}{\Lambda_{\sigma}}^a]\notag\\& +B^{a}{}_{0\mu}{\Lambda_{\nu}}^{0}{\Sigma_\alpha}^{\rho}+B^{a}{}_{0\nu}{\Lambda_{\mu}}^0{\Sigma_a}^{\rho}+B^{a}{}_{\mu b}{\Lambda_{\nu}}^b{\Sigma_\alpha}^{\rho}+B^{a}{}_{\nu b}{\Lambda_{\mu}}^b{\Sigma_a}^{\rho}
\label{comid}
\end{align}
For the standard N-C geometry, torsion vanishes. This means the connection $
\Gamma_{\nu\mu}^{\rho}$ is symmetric. Assuming that the connection is symmetric,
\begin{equation}
\Gamma_{\nu\mu}^{\rho}-\Gamma_{\mu\nu}^{\rho}=0
\end{equation}
To get the desired expression for the symmetric connection we need a little more algebra. 
We can write from the first line of (\ref{PP}),
\begin{equation}
\partial_{\mu}{\Lambda_{\nu}}^a {\Sigma_a}^{\rho}-\partial_{\nu}{\Lambda_{\mu}}^a {\Sigma_a}^{\rho}+B^{a}{}_{\mu b}{\Lambda_{\nu}}^b{\Sigma_a}^{\rho}-B^{a}{}_{\nu b}{\Lambda_{\mu}}^b{\Sigma_a}^{\rho}=0
\label{con}
\end{equation}
After a little calculation we can get,
\begin{equation}
\frac{1}{2}h^{\rho\sigma}[-{\Lambda_{\nu}{}^a}\partial_{\mu}{\Lambda_{\sigma}}^a
-{\Lambda_{\mu}{}^a}\partial_{\nu}{\Lambda_{\sigma}}^a]=-\partial_{\sigma}h_{\mu\nu}-B^{a}{}_{\mu b}{\Lambda_{\nu}}^b{\Sigma_a}^{\rho}-B^{a}{}_{\nu b}{\Lambda_{\mu}}^b{\Sigma_a}^{\rho}
\label{conec}
\end{equation}
Using (\ref{conec}) we can get the cherished form of the connection from (\ref{comid}) that reproduces the structure (\ref{covariantconnection}),
where the two form K is defined as,
\begin{align}
 h^{\rho\lambda}K_{\lambda(\mu}\tau_{\nu)} &=\frac{1}{2}h^{\rho\lambda}[K_{\lambda\mu}\tau_{\nu}+K_{\lambda\nu}\tau_{\mu}]\notag\\
 &=\frac{1}{2}[B^{a}{}_{0\mu}{\Lambda_{\nu}}^{0}{\Sigma_\alpha}^{\rho}+B^{a}{}_{0\nu}{\Lambda_{\mu}}^0{\Sigma_a}^{\rho}]
\end{align}
The construction of the N-C geometry in the standard form is now complete.

\subsubsection{Horava-Lifshitz geometry}

We have provided an algorithm (see section 3.1) to convert a globally Galileo symmetric model (\ref{globalaction})
to one that is symmetric under local Galileo transformations (\ref{localgalilean}). The structure of this new model is presented in (\ref{localschaction}).
The localization process of Galilean symmetry requires the introduction of a set of geometric fields that can be arranged as a $4 \times 4$ matrix (\ref{Smatrix1}). In the previous section we have used these fields to get Newton Cartan geometry, following Cartan's prescription of a timelike vector with a nondegenerate $3\times 3$ kernel constituting the degenerate metric structure. In this section we show that a different non relativistic geometry can be obtained using the fields contained in (\ref{Smatrix1}) which can be identified with the projectable version of the Horava Lifshitz geometry.

Recall that the elements of the matrices (\ref{Smatrix1}) and its inverse converted the global covariant derivatives to their local counterparts and conversely, during the gauging process. As remarked earlier they carry two indices: the local index $\alpha (\bar{0}, a)$ and the global index $\mu (0,i)$. The transformations (\ref{delth3}) and (\ref{delth4}) are like a co (contra)variant vector under the foliation preserving diffeomorphism (\ref{localgalilean}) corresponding to a global index. The local indices characterise Galilean boost or rotation in flat eucledian space and universal time. The matrices $\Sigma_\alpha^{{}\mu}$ (or its inverse) can thus be identified with the vielbeins
that link the local basis in the tangent space with the global coordinate basis. In the last section we have seen how the Newton-Cartan geometry can be obtained from these geometric elements. Here we will show that a nonsingular metric can be constructed which gives the projectable version of the Horava gravity. 

  The geometric interpretation of the Galilean gauge theory brings us close to the space time of Horava gravity. Before we introduce more geometric objects it is useful to note an important point. In the locally Minkowskian tangent space both boost and rotations are transformations that keep the origin of the coordinate system invariant. But in the Galilean space time no natural space time metric exists and the boosts affect time and space asymmetrically. If we keep this asymmetry, a single non degenerate spacetime metric (which is necessarily symmetric) does not exist \cite{Cartan-1923, Cartan-1924}. This line of analysis leads to the Newton - Cartan spacetime which is characterised by two degenerate metrices eventually providing the geometric formulation of Newton's gravity. On the other hand a single nondegenerate metric must be assumed for the Horava geometry. As the esuing analysis shows it is possible to achieve such a metric by putting the boost parameter vanishing. 
   Such a choice immediately leads to Horava's construction as will be demonstrated below.

  We now propose the following `metric'
\begin{equation}
g_{\mu\nu} = \eta_{\alpha\beta}\Lambda_\mu^{{}\alpha}\Lambda_\nu^{{}\beta}\label{metric}
\end{equation}
and work out its variation under (\ref{localgalilean}). Both $g_{00}$ and $g_{ij}$ transform as second rank tensors,
\begin{eqnarray}
\delta_0 g_{00} &=& -\xi^\nu\partial_\nu g_{00} - \partial_0\xi^\rho g_{\rho 0} - \partial_0\xi^\rho g_{0 \rho} \nonumber\\
\delta_0 g_{ij} &=& -\xi^\nu\partial_\nu g_{ij} - \partial_i\xi^\rho g_{\rho j} - \partial_j\xi^\rho g_{i \rho}\label{mettrans}
\end{eqnarray}
In the above deduction the definition (\ref{metric}) and the transformations(\ref{delth3}) and (\ref{delth4}) 
 are used. The variation of $g_{0k}$ comes out as
\begin{equation}
\delta_0 g_{0j} = -\xi^\nu\partial_\nu g_{0j} - \partial_0\xi^\rho g_{\rho j} - \partial_j\xi^\rho g_{0 \rho}+ v_b\Lambda_0^{\bar{0}}\Lambda_k^{{}b}\label{mettrans1}
\end{equation}
The term containing the boost parameter will be dropped following our previous argument. 
Thus the coefficients $g_{0j}$ also satisfy the required tensorial transformation properties.

The significance of the 'metric' is to be understood properly. The coordinates $x^\mu$ define a four dimensional differentiable manifold whose physical structure is $R^3\times R$. The 'metric' imposes a Riemannian structure on this manifold. This metric is constructed from the vielbeins arising out of our localisation procedure. It is nonsingular and symmetric. The inverse of $g_{\mu\nu}$ can be easily constructed as
\begin{equation}
^{(4)}g^{\mu\nu} = \eta^{\alpha\beta}{\Sigma_\alpha}^{{}\mu}{\Sigma_\beta}^{{}\nu}\label{metricinv}
\end{equation}
The invariant 'interval' corresponding to it is not the same as that of the Galileo - Newton space time. But it helps us to implement a foliation through the Arnowit - Deser - Misner (ADM) construction in general relativity. So we define the lapse N and the shift variables in the usual way 
\begin{eqnarray}
N &=&\left(-g^{00}\right)^{1/2}\nonumber\\
N^j&=&g^{ij}g_{0i}
\end{eqnarray}
where $g^{ij}$ is the inverse of the spatial part of the metric
$g_{\mu\nu}$

We recall that the physical variables of the Horava gravity are $N, N_i$ and $g_{ij}$ which have definite transformation laws. These laws are easily calculated from the above relations,
\begin{align}
\delta g_{ij}&=-\partial_i\xi^k g_{jk}-\partial_j\xi^kg_{ik}-\xi^k
\partial_k g_{ij}-\xi^0{\dot {g}}_{ij},\nonumber\\
\delta N_i&=-\partial_i\xi^jN_j-\xi^j\partial_jN_i-\dot\xi^jg_{ij}-\dot{\xi}^0 N_i-\xi^0
\dot {N}_i,\nonumber\\
\delta N &=-\xi^j\partial_j N-\dot{\xi}^0 N-\xi^0\dot{N}.
\label{foldif}
\end{align}
They are exactly the same as found by taking the $c\to \infty$ limit of the ADM decomposition of the metric in general relativity, which is the prescription followed by Horava.

 Let us pause to think what we have achieved. We have reinterpreted the Galilean gauge theory in flat Euclidean space with absolute time as a geometric theory over a curved manifold. It is a differentiable manifold which is left invariant by the foliation preserving diffeomorphism (\ref{localgalilean}). The space is converted to a metric space by constructing a metric which has all its desired properties, namely, it is a second rank covariant tensor under $Diff_F \hskip,2cm \rm{on} \hskip.2cm {\cal{M}}$, nonsingular and symmetric. An ADM splitting of this manifold exists and as usual the physical variables are identified as $g_{ij}$, the spatial part of the metric, $N$, the lapse and $N^i$, the shift variables. The transformation rules of these variables are given by (\ref{foldif}). These are the same transformations obtained in \cite{H}. Naturally we would like to identify the space time given by the metric (\ref{metric}) with that of Horava - Lifshitz gravity. But one piece of the dictionary is still void. It is the invariant measure of the volume. From Galilean gauge theory we have identified the measure as (see equation (\ref{measure}))$ 
 \int dt d^3x \frac{M}{\theta}$. From the definitions of $M$ and $\theta$ and that of the metric (\ref{metric}) we find 
that
\begin{equation}
 \int dt d^3x \frac{M}{\theta} = \int dt d^3x \sqrt{\det{g_{ij}}}N
\end{equation}
which reproduces the measure of Horava-Lishitz gravity.

\section{Concluding remarks}
This paper is a comprehensive account of the recently discovered theory of localising the extended Galilean symmetry and extracting nonrelativistic diffeomorphism therefrom \cite{BMM1, BMM2, BMM3}, in both spatial and space time manifold. Our theory is inspired by Utiyama's approach of gauging the Poincare symmetry \cite{Utiyama:1956sy} --\cite{ sciama} that led to the emergence of Riemann-Cartan space-time. In the relativistic case, the geometrical structures are very well known. There exists a fully geometrical method due to Cartan which treats the symmetries in the tangent space. Vielbeins are introduced which amount to a factorisation of the metric. In fact
Newtonian gravity was also cast as a geometric theory on 4-dimensional spacetime by Cartan \cite{Cartan-1923,Cartan-1924}, and was subsequently developed by many stalwarts \cite{Havas} to \cite{MALA}. But the peculiarity of the Galileo - Newton relative space and universal time, manifested in the degenerate metric structure of Newton-Cartan geometry, makes the gauging of Galilean symmetry of a field theory far more intricate and subtle. Barring a single paper (to the best of our knowledge) \cite{PLP} there was no attempt to follow Utiyama's approach. This solitary example is again based on particle mechanics and did not try to connect with geometry.

After several studies \cite{BMM1}, \cite{BMM2}, \cite{BMM3} the idea of gauging the non-relativistic (extended Galilean) symmetry{\footnote{Note that we are distinguishing between gauging of the extended Galilean symmetry and extended Galilean (Bargmann)algebra. As we have emphasised, by a gauging of the symmetry we imply that the original global parameters of the symmetry  are localised. The implications of this for discussing non relativistic diffeomorphism invariance has been considered here. On the other hand gauging the Bargmann algebra, which is the more popular route for discussing these issues , has been considered by other authors , for instance \cite{B}}} initiated by us in \cite{BMM1} has eventually taken a concrete and definite shape.
We therefore felt the need of a thorough and systematic exposure of our method. Many old results have been presented in a new light with fresh insights. To facilitate the exposition of our work a short review of gauging Poincare algebra is provided where the techniques of nonabelian gauge theory have been used. This is compared with the method of gauging the symmetry that has been adopted here. In the  algebraic approach, the first step in establishing a correspondence with diffeomorphism symmetry requires a vanishing curvature. This implies a torsion free theory. Apart from this a geometrical input is necessary to complete the correspondence. Our analysis  of the problem therefore clearly shows that algebraic techniques alone are not sufficient to connect with (even) the torsion free Riemannian geometry. This is contrasted with the approach of gauging the Poincare symmetry of the action of a generic field theory in the Minkowski space. We find the structure of Riemann - Cartan spacetime to emerge seamlessly. This shows the power and generality of our scheme. 

We have given entirely new results corresponding to spatial and space time nonrelativistic diffeomorphism invariance, concerning the coupling of Schr\"odinger field, which is a complex scalar in nature, with the spin connection of the curved two dimensional space. This mechanism was shown to lead to the appearance of Chern - Simons dynamics and the Wen - Zee term in the theory of fractional quantum Hall effect \cite{F}. We have also demonstrated in this paper that the space time nonrelativistic diffeomorphism invariance, obtained in our method, is poignant with new possibilities. Earlier, following the lead in Cartan's reduction of the Newton Cartan metric in a temporal flow along with its nondegenerate kernel, we divided the `vielbein matrix' analogously. The outcome was Newton Cartan geometry. In this paper we have shown that if the whole `vielbein matrix' is used to define a single nondegenerate metric, the result is Horava geometry. A remarkable observation is
that Galilean boost is no longer a symmetry in the tangent space of Horava geometry. This exhibits its difference with Newton Cartan geometry. However, both have a common origin that is manifested here in the interpretation of the"vielbein matrix". Indeed there should be a common origin since these are theories of non-relativistic gravity.

To put our work in a proper perspective  one has to follow how interest in the problem of nonrelativistic diffeomorphism has been rekindled in the recent past. In their study of geometric properties of the Hall fluid in the lowest Landau level, Son and Wingate used the idea of coupling a Galilee symmetric theory in curved space \cite{SW,HS}. Later, the approach was used by \cite{M1, M2,Son,F} to name a few. The empirical approach of \cite{SW} raised many questions. These questions could not be answered satisfactorily in an otherwise very rich literature on Galilean symmetry. A spate of papers appeared \cite{Son, Gromov, B} but none could provide an algorithmic approach that was needed. In \cite{B} the procedure of gauging the Bargmann algebra was performed. This could give a truncated vielbein approach but devoid as it is of any dynamical background, the results were of restricted utility. Our method, as has been elaborated in this paper, eradicates all these shortcomings. Meanwhile, nonrelativistic diffeomorphism invariance has found possible applications to NR superparticle theory \cite{BGJKM}, fractional quantum Hall effect \cite{SW} and also in holographic theories \cite{Wu}. Thus its importance in the present context, cannot be overemphasized.

      
A general lagrangian field theory invariant under the Galilean group of transformations has been considered. The parameters of this group of transformations are constants. Localisation of these transformations means that the parameters become functions of space and time. Naturally, due to the presence of derivatives, the original theory is no longer invariant under these local galilean transformations. The invariance is brought back in two steps. First, locally covariant derivatives were obtained that transformed under the local transformations in the same way as the partial derivatives under global transformations. Secondly, the integration measure was required to be suitably changed. The algorithm was illustrated by taking a nonrelativistic (Schr\"odinger) field theory as an example.
    
    The inclusion of the Chern - Simons action in nonrelativistic curved space has been profusely discussed in the literature \cite{HS}, \cite{F}. To see how this could be implemented in our formalism we first showed  how  gauge fields can be  naturally accommodated. The inclusion of Chern- Simons term was then considered. The whole machinery was used to develop a geometrical interpretation of our constructions. Spatial nonrelativistic diffeomorphism invariance, as comprehensively obtained in this paper, is sure to equip the condensed matter theorist with new tools for its perusal. Particlar emphasis has been given to the Chern-Simons dynamics and new results are found which clarify the coupling of the scalar fields with the spin connection in two space dimensions -- a result at present understood in quite an involved manner \cite{F}. Our analysis gives a conceptually clean picture. The structure of the covariant derivative found here - a consequence of the coupling - is known \cite{F} to yield the Hall viscosity and Wen-Zee term in the discussion of FQHE.
 
For vanishing time translation parameter, our algorithm was able to reformulate nonrelativistic field theories in Euclidean space and universal time, invariant under local rotation and  boosts, to a field theory in curved space. This means that a general prescription was obtained to formulate a nonrelativistic diffeomorphism invariant theory. As already mentioned such theories find applications in various contexts, especially in the study of fractional quantum Hall effect. 
 
 Our method is also successful in achieving spatial NRDI in the context of fluid model. This model is a hydrodynamic form of the Schr\"odinger field theory considered here. It may be noted that in reference \cite{BJ} an attempt was made to discuss diffeomorphism inavariance in this model. However one had to invoke field dependent diffeomorphism. We have, on the contrary, shown that it is possible to discuss the conventional diffeomorphism symmetry by a  systematic extension of the model where ordinary derivatives have been replaced by covariant derivative and an appropriate correction in the measure has been incorporated.
 
The geometrical content of our theory is certainly not confined to nonrelativistic diffeomorphism invariance in space. We pushed it forward in a big way. Introducing a 4-dimensional manifold we were able to identify a ($4\times 4$) invertible matrix structure, the vielbein fields, from the set of fields introduced during gauging. Using only the vielbein postulate we were able to endow the  4-dimensional manifold with structures that makes it equivalent to the Newton-Cartan spacetime. It was indeed gratifying to observe how the transformation rules obtained during the localisation procedure provided the correct geometrical transformations to the various objects of the Newton-Cartan spacetime. This was indicative of the internal consistency of the algorithm.

The NRDI that we have formulated in Galileo Newton space time allows new possibilities. Besides the ones that have already been discussed, we have provided a geometric basis for the construction of Horava - Lifshitz theory of gravity which was unclear in its original formulation \cite{H}. The genesis of the foliation preserving diffeomorphism invariant space time of Horava is shown to originate from the localisation of non-relativistic symmetry subject to a particular condition. This condition is the vanishing of the boost parameter. This is done on the ground that in the non-relativistic case, there is no single nondegenerate space time metric. Indeed, if we had retained the boost parameter it would have led to the Newton - Cartan space time as shown in section 4.7. The Newton - Cartan space time, as is well known, is the basis for the construction of Newton's gravity as a space time phenomenon. This clearly shows the difference between the geometric aspects of Newton's formulation vis-a-vis Horava's formulation. However, we must note the common origin of both these types of non-relativistic gravity. The non-relativistic diffeomorphism associated with these theories emanate from a gauging of the non-relativistic space time symmetries. Retaining the boost parameter would lead to Newton's gravity, while setting it to zero leads to Horava - Lifshitz formulation.

\end{document}